\documentclass[prd,letterpaper,nofootinbib,twocolumn,superscriptaddress,aps,showpacs]{revtex4}
\usepackage[dvipsnames]{xcolor}
\usepackage{fancyhdr}
\usepackage{float}
\usepackage{listings}
\usepackage{enumerate}
\usepackage[T1]{fontenc}
\usepackage{ae}
\usepackage{rotating}
\usepackage{subfigure}
\usepackage{amsmath}
\usepackage{amssymb}
\usepackage{graphicx}
\usepackage{dcolumn}
\usepackage{bm}
\usepackage{hyperref}
\usepackage{morefloats}
\usepackage{slashed}
\usepackage{dcolumn}
\usepackage{tabularx}
\usepackage[sort&compress]{natbib}
\usepackage[normalem]{ulem}

\newcommand{\msout}[1]{\text{\sout{\ensuremath{#1}}}}

\newcommand{\ibardd}{\ddot{\msout{I}}_{ij}}
\newcommand{\ibarddxx}{\ddot{\msout{I}}_{xx}}
\newcommand{\ibarddyy}{\ddot{\msout{I}}_{yy}}
\newcommand{\ibarddzz}{\ddot{\msout{I}}_{zz}}
\newcommand{\ibarddxy}{\ddot{\msout{I}}_{xy}}
\newcommand{\ibarddxz}{\ddot{\msout{I}}_{xz}}
\newcommand{\ibarddyz}{\ddot{\msout{I}}_{yz}}

\newcommand{\ibarddtilde}{\tilde{\ddot{\msout{I}}}_{ij}}

\newcommand{\egw}{E_{\mathrm{GW}}}
\newcommand{\dedf}{\frac{\mathrm{d}\egw}{\mathrm{d}f}}
\newcommand{\dedflong}{\mathrm{d}\egw/\mathrm{d}f}
\newcommand{\fpeak}{f_{\mathrm{peak}}}

\newcommand{\hrss}{h_{\mathrm{rss}}}
\newcommand{\hp}{h_{+}}
\newcommand{\hc}{h_{\times}}
\newcommand{\hlm}{H_{lm}(t)}
\newcommand{\sylm}{{}^{-2}Y_{lm}(\iota,\phi)}
\newcommand{\htz}{H_{20}^{\mathrm{quad}}}
\newcommand{\hto}{H_{2\pm1}^{\mathrm{quad}}}
\newcommand{\htt}{H_{2\pm2}^{\mathrm{quad}}}
\newcommand{\ytz}{{}^{-2}Y_{20}}
\newcommand{\yto}{{}^{-2}Y_{2\pm1}}
\newcommand{\ytt}{{}^{-2}Y_{2\pm2}}

\newcommand{\hrssf}{h_{\mathrm{rss}}^{50\%}}

\newcommand{\Msun}{M_{\odot}}
\newcommand{\Rsun}{R_{\odot}}

\newcolumntype{d}[1]{D{.}{.}{#1}}
\newcommand\ligodoc{P1400233}

\begin{document}

\title{Observing gravitational waves from core-collapse supernovae in the\\
advanced detector era}

\newcommand*{\ligoCIT}{LIGO - California Institute of Technology, Pasadena, CA  91125, USA}
\affiliation{\ligoCIT}

\newcommand*{\CaRT}{TAPIR, MC 350-17, California Institute of Technology, Pasadena, CA  91125, USA}
\affiliation{\CaRT}

\newcommand*{\ER}{Embry Riddle University, 3700 Willow Creek Road, Prescott, AZ, 86301, USA}
\affiliation{\ER}

\newcommand*{\Cardiff}{Cardiff University, Cardiff, CF24 3AA, United Kingdom}
\affiliation{\Cardiff}

\newcommand*{\LLO}{LIGO Livingston Observatory, Livingston, LA 70754, USA}
\affiliation{\LLO}

\newcommand*{\LSU}{Louisiana State University, Baton Rouge, LA 70803, USA}
\affiliation{\LSU}


\author{S. E.~Gossan} \affiliation{\ligoCIT}\affiliation{\CaRT}

\author{P. Sutton} \affiliation{\Cardiff}

\author{A. Stuver} \affiliation{\LLO}\affiliation{\LSU}

\author{M. Zanolin} \affiliation{\ER}

\author{K. Gill} \affiliation{\ER}

\author{C. D.~Ott} \affiliation{\CaRT}

\date{\today}

\newcommand{\sg}[1]{\textcolor{magenta}{#1}}

\begin{abstract}
The next galactic core-collapse supernova (CCSN) has already exploded, and its
electromagnetic (EM) waves, neutrinos, and gravitational waves (GWs) may arrive
at any moment. We present an extensive study on the potential sensitivity of
prospective detection scenarios for GWs from CCSNe within $5\,\mathrm{Mpc}$, 
using realistic noise at the predicted sensitivity of the Advanced LIGO and 
Advanced Virgo detectors for 2015, 2017, and 2019. We quantify the 
detectability of GWs from CCSNe within the Milky Way and Large Magellanic Cloud, 
for which there will be an observed neutrino burst. We also consider extreme 
GW emission scenarios for more distant CCSNe with an associated EM signature. 
We find that a three-detector network at design sensitivity will be able to detect
neutrino-driven CCSN explosions out to $\sim5.5\,\mathrm{kpc}$, while rapidly 
rotating core collapse will be detectable out to the Large Magellanic Cloud at 
$50\,\mathrm{kpc}$. Of the phenomenological models for extreme GW emission 
scenarios considered in this study, such as long-lived bar-mode instabilities 
and disk fragmentation instabilities, all models considered will be detectable 
out to M31 at $0.77\,\mathrm{Mpc}$, while the most extreme models will be detectable 
out to M82 at $3.52\,\mathrm{Mpc}$ and beyond.
\end{abstract}


\maketitle

\section{Introduction}
Core-collapse supernovae (CCSNe) are driven by the release of
gravitational energy in the core collapse of massive stars in the
zero-age-main-sequence mass range $8\,M_\odot \lesssim M \lesssim
130\,M_\odot$. The available energy reservoir of $\sim$$300$~Bethe
(B, $1\, \mathrm{B} = 10^{51}\,\mathrm{erg}$) is set by the difference
in gravitational binding energy of the precollapse core ($R\sim
1000$-$2000\,\mathrm{km}$, $M \sim 1.4\,M_\odot$) and the collapsed
remnant ($R\sim10$-$15\,\mathrm{km}$). Much of this energy is initially
stored as heat in the protoneutron star and most of it ($\sim$99\%) is
released in the form of neutrinos, $\sim 1\%$ goes into the kinetic
energy of the explosion, $\sim0.01\%$ is emitted across the
electromagnetic (EM) spectrum, and an uncertain, though likely smaller,
fraction will be emitted in gravitational waves 
(GWs)~\cite{ott:09,kotake:13review}.

Distant CCSNe are discovered on a daily basis by astronomers. Neutrinos from
CCSNe have been observed once, from the most recent nearby CCSN, 
SN 1987A~\cite{hirata:87,bionta:87}, 
which occured in the Large Magellanic Cloud (LMC), roughly $52\,\mathrm{kpc}$ from 
Earth~\cite{panagia:99}. GWs\footnote{For detailed reviews of GW theory and 
observation, we refer the reader to 
Refs.~\cite{mtw,thorne:87,flanagan:05,creighton:11book}.}
are --at lowest and likely dominant order-- emitted by quadrupole
mass-energy dynamics. In the general theory of relativity, GWs have two 
polarizations, denoted plus $(+)$ and cross $(\times)$. Passing GWs
will lead to displacements of test masses that are directly proportional to the
amplitudes of the waves and, unlike EM emission, not their
intensity. GWs have not yet been directly detected.

GWs, much like neutrinos, are emitted from the
innermost region (the core) of the CCSN and thus convey information on 
the dynamics in the supernova core to the observer. They
potentially carry information not only on the general degree of
asymmetry in the dynamics of the CCSN, but also more directly on the
explosion mechanism~\cite{ott:09,ott:09b,logue:12}, on the structural
and compositional evolution of the protoneutron 
star~\cite{murphy:09,mueller:13gw,yakunin:10,yakunin:15}, the rotation rate of the
collapsed core~\cite{summerscales:08,dimmelmeier:08,hayama:09,abdikamalov:10}, and
the nuclear equation of state~\cite{roever:09,marek:09b,dimmelmeier:08}.

A spherically symmetric CCSN will not emit GWs.
However, EM observations suggest
that many, if not most, CCSN explosions exhibit asymmetric features
(e.g.,~\cite{wang:08,chornock:11,smith:12a,sinnott:13,boggs:15}). This is also 
suggested by results of multidimensional CCSN simulations 
(e.g.,~\cite{kotake:12snreview,janka:12a,ott:13a,dolence:13,hanke:13,moesta:14,couch:14,lentz:15,melson:15} 
and references therein). Spherical symmetry should be robustly broken by stellar
rotation, convection in the protoneutron star and in the region behind
the CCSN shock, and by the standing accretion shock instability (SASI~\cite{blondin:03}). 
The magnitude and time variation of
deviations from spherical symmetry, and thus the strength of the
emitted GW signal, are uncertain and likely vary from
event to event~\cite{mueller:13gw,ott:09}. State-of-the-art models,
building upon an extensive body of theoretical work on the
GW signature of CCSNe, predict GW strains--relative displacements of test masses in
a detector on Earth--$h$ of order $10^{-23}$-$10^{-20}$ for a core
collapse event at $10\,\mathrm{kpc}$, signal durations of
$1\,\mathrm{ms}$-$\mathrm{few}\,\mathrm{s}$, frequencies of
$\sim$$1$-$\,\mathrm{few}\,1000\,\mathrm{Hz}$, and total emitted
energies $E_\mathrm{GW}$ of $10^{41}$-$10^{47}\,\mathrm{erg}$
(corresponding to $10^{-12}$-$10^{-7}\,M_\odot c^2$)~\cite{ott:09,kotake:12snreview,
mueller:13gw,ott:13a,mueller:e12,kuroda:14,yakunin:10,dimmelmeier:08,ott:12a,ott:11a}.
More extreme phenomenological
models, such as long-lasting rotational instabilities of the proto-neutron 
star and accretion disk fragmentation instabilities, associated with 
hypernovae and collapsars 
suggest much larger strains and more energetic emission, with
$E_\mathrm{GW}$ perhaps up to $10^{52}\,\mathrm{erg}$
($\sim0.01\,M_{\odot}\,c^2$)~\cite{fryer:02,vanputten:04,piro:07,corsi:09}.

Attempts to detect GWs from astrophysical sources were
spearheaded by Weber in the 1960s~\cite{weber:60}. Weber's detectors
and other experiments until the early 2000s relied primarily on
narrow-band ($\lesssim 10$s of Hz) resonant bar or sphere 
detectors (e.g.,~\cite{aufmuth:05}). 
Of these, NAUTILUS~\cite{astone:13bar},
AURIGA~\cite{vinante:06auriga}, and Schenberg~\cite{aguiar:12schenberg} are 
still active. The era of broadband
GW detectors began with the kilometer-scale
first-generation laser interferometer experiments. The two 4-km 
LIGO observatories~\cite{ligo:09} are in Hanford, Washington and Livingston,
Louisiana, hereafter referred to as H1 and L1, respectively. A second
2-km detector was located in Hanford, referred to as H2, but was
decommissioned at the end of the initial LIGO observing runs.
The 3-km Virgo detector~\cite{ivirgo:12} is located in
Cascina, Italy. Other GW interferometers are
the 300-m detector TAMA300~\cite{tama:01} in Mitaka, Japan, and the
600-m detector GEO600~\cite{grote:10} in Hanover, Germany.
The second generation 
of ground-based laser interferometric GW detectors, roughly 10 times
more sensitive than the first generation, are under construction. The two
Advanced LIGO detectors~\cite{harry:10} began operation in late 2015 at
approximately one-third of their final design sensitivity, jointly with
GEO-HF~\cite{geohf:06}. Advanced Virgo~\cite{advirgo} will commence
operations in 2016, followed by KAGRA~\cite{kagra:12} later in the
decade. LIGO India~\cite{ligoindia:11} is under consideration, and may begin
operations c. 2022.

Typically, searches for GW transients must scan the entire GW detector
data set for signals incident from any direction on the sky
(e.g.,~\cite{abadie:12s6burst,abadie_s5y2_burst:10} and references therein) 
unless an external ``trigger" is available. The observation of an 
EM or neutrino counterpart can provide timing and/or sky position
information to localize the prospective GW signal
(e.g.,~\cite{abadie:12s6grb,ligo_051103:12,ligo_a5_sgr:11} and references
therein). The sensitivity of GW searches utilizing external triggers can 
be more sensitive by up to a factor of $\sim2$, as constraints
on time and sky position help reduce the background noise present in
interferometer data (e.g.,~\cite{abadie:12s6grb,was:12}). In both cases, 
networks of two or more
detectors are typically required to exclude instrumental and local environmental
noise transients that could be misindentified as GW
signals. This is particularly
important in the case where there is no reliable model for the GW signal, such
as for CCSNe.

Arnaud~\emph{et al.}~\cite{arnaud:99} were the
first to make quantitative estimates on the detection of GWs from 
CCSNe. They studied the detectability of GW signals from axisymmetric rotating 
core collapse~\cite{zwerger:97}, by means of three different filtering
techniques. The authors showed that, in the context of stationary, Gaussian noise with
zero-mean, the signals should be detectable throughout the galaxy with initial
Virgo~\cite{ivirgo:12}.

Ando~\emph{et al.}~\cite{ando:05tama}, using single-detector
data taken with the TAMA300 interferometer, were the first to carry
out an untriggered all-sky blind search specifically for GWs
from rotating core collapse. These authors employed a
model-independent approach which searches for time-frequency regions
with excess power compared to the noise background (called an
``excess power method" (e.g.,~
\cite{anderson:99,anderson:01,vicere:02,sylvestre:03}). They employed rotating
core-collapse waveforms from Dimmelmeier~\emph{et al.}~\cite{dimmelmeier:02} 
to place upper limits on detectability and rate
of core collapse events in the Milky Way. Unfortunately, these upper limits 
were not astrophysically interesting due to the high false alarm rate of
their search, caused by their single-detector analysis and the limited
sensitivity of their instrument.

Hayama~\emph{et al.}~\cite{hayama:15} studied the detectability of
GWs from multidimensional CCSN simulations 
from~\cite{kuroda:14,kotake:09,takiwaki:11,kotake:11}. Using the coherent
network analysis network pipeline \texttt{RIDGE}~\cite{hayama:07}, signals in 
simulated Gaussian noise for a four-detector network
containing the two Advanced LIGO detectors, Advanced Virgo, and KAGRA are
considered. The authors find that GWs from the neutrino-driven explosions
considered are detectable out to
$\sim$(2-6)\,kpc, while GWs from rapidly rotating core-collapse and
nonaxisymmetric instabilities are detectable out to between 
$\sim$(11-200)\,kpc.

In this article, we describe a method for the detection of
GWs from CCSNe in nonstationary, non-Gaussian data recolored to the 
predicted sensitivity of the second-generation interferometers. Since GW 
emission from CCSNe may be very weak (but can vary by orders of magnitudes in
strain, frequency content, and duration), we follow a triggered
approach and employ \texttt{X-Pipeline}~\cite{sutton:10}, a coherent
analysis pipeline designed specifically to detect generic GW transients
associated with astrophysical events such as gamma-ray bursts and supernovae
using data from networks of interferometers.
We consider 
\begin{enumerate}
\item CCSNe within $\sim50\,\mathrm{kpc}$ with
sky position and timing localization information provided by neutrinos 
(e.g.,~\cite{beacom:99a,pagliaroli:09,muehlbeier:13}). At close source
distances, we hope to detect GWs from CCSNe in current scenarios 
predicted by state-of-the-art multidimensional numerical simulations.
\item Distant CCSNe with sky position and timing 
localization information provided by EM observations. At distances 
greater than $\sim$ (50-100)\,kpc, we do not
expect to detect GWs from the conservative emission scenarios
predicted by multidimensional CCSN simulations. Instead, we consider
more extreme, phenomenological emission models. These may be unlikely
to occur, but have not yet been constrained observationally.
\end{enumerate}
We consider GW emission from ``garden-variety" CCSNe 
(e.g.~convection, SASI, and rotating core collapse and bounce) with 
waveform predictions from multidimensional CCSN simulations, in 
addition to extreme postcollapse GW emission mechanisms. In addition, we 
consider for both scenarios sine-Gaussian GW bursts as an 
\emph{ad hoc} model for GW signals of central frequency $f_{0}$ and quality factor $Q$, 
which are frequently used to assess the sensitivity of searches
for generic GW bursts of unknown morphological
shape~\cite{abadie_s5y2_burst:10,abadie:12s6burst}.

This paper is organized as follows. In Sec.~\ref{sec:challenges}, we discuss 
the challenges associated with observing GWs from CCSNe. We outline our
strategies to overcome these challenges and introduce the observational scenarios
considered in this study in Sec.~\ref{sec:obs_scenarios}. We 
review the waveforms from multidimensional hydrodynamic simulations and 
phenomenological waveform models used in this study in Sec.~\ref{sec:models}. In
Sec.~\ref{sec:GWdetection}, we give details of our analysis approach
and lay out how we establish upper limits for
detectability. We present the results of our analysis and provide
quantitative estimates for the distances out to which GWs may be observed
for each of the considered waveform models and detector sensitivity in
Sec.~\ref{sec:results}. We summarize and conclude in
Sec.~\ref{sec:conclusions}.

\section{Challenges}
\label{sec:challenges}

GW astronomers looking for short-duration GW transients emitted from CCSNe
face multiple challenges.

\subsection{The rate of observable events is low}
If GW emission in standard, ``garden-variety" CCSNe occurs at the strains 
and frequencies predicted by current models, simple estimates of signal-to-noise 
ratios (SNRs) suggest that even second-generation detectors may be limited to 
detecting core-collapse events in the Milky Way and the Small and Large Magellanic 
Clouds~\cite{ott:09,ott:13a,yakunin:10,murphy:09}. The expected rate of CCSNe in the 
Milky Way is $\sim$ (0.6-10.5)$\times10^{-2}\,\mathrm{CCSNe}\,\mathrm{yr}^{-1}$, 
(e.g.,~\cite{vandenbergh:91,cappellaro:93,tammann:94,li:11b,diehl:06,adams:13}),
and it is 
$\sim$ (1.9-4.0)$\times10^{-3}\,\mathrm{CCSNe}\,\mathrm{yr}^{-1}$ in the 
combined Magellanic Clouds~\cite{tammann:94,maoz:10,vandenbergh:91}. 

Similar SNR estimates for extreme GW emission models for CCSNe suggest that
they may be observable throughout the Local Group and beyond ($D \lesssim
20\,\mathrm{Mpc}$). Within the local group ($D \lesssim 3\,\mathrm{Mpc}$), the
CCSN rate is $\sim9\times10^{-2}\,\mathrm{CCSNe}\,\mathrm{yr}^{-1}$, with
major contributions from Andromeda (M31),
Triangulum (M33), and the dwarf irregular galaxy IC 10, IC 1613, and NGC
6822~\cite{timmes:97a,thronson:90,tammann:94,vandenbergh:91}. Outside
of the Local Group, the CCSN rate increases to
$\sim0.15\,\mathrm{CCSNe}\,\mathrm{yr}^{-1}$ within $D \sim 5\,\mathrm{Mpc}$,
including IC 342, the M81 group, M83, and NGC 253 as significant contributors to
the CCSN rate~\cite{gill:15,ando:05,kistler:13,kistler:11,botticella:11,mattila:12}. 
Within $D = 10\,\mathrm{Mpc}$, the CCSN rate is 
$\sim 0.47\,\mathrm{CCSNe}\,\mathrm{yr}^{-1}$, 
while it increases to $\sim2.1\,\mathrm{CCSNe}\,\mathrm{yr}^{-1}$ 
within $D =
20\,\mathrm{Mpc}$~\cite{botticella:11,mattila:12,kistler:11,gill:15}.

\subsection{The duty cycle of the detectors is not 100\%}
\label{subsec:dutycycle}
The fraction of time interferometers are operating and taking
science-quality data is limited by several factors including commissioning
work (to improve sensitivity and stability) and interference
due to excessive environmental noise.

For example, consider LIGO's fifth science run (S5), the data from which we use
for the studies in this paper. S5 lasted almost two years between
November 15 2005 through November 02 2007, and 
the H1, H2, and L1 detectors had duty cycles of $75\%$, $76\%$, and $65\%$,
respectively. The duty cycle for double coincidence (two or more detectors
taking data simultaneously) was $60\%$, and the triple coincidence duty
cycle was $54\%$~\cite{abbott:09_S5burst,abadie:10_s5calib}. The risk of completely
missing a CCSN GW signal is mitigated by having a larger network of
detectors. In addition, resonant bar and sphere detectors do provide
limited backup~\cite{astone:13bar,vinante:06auriga,aguiar:12schenberg}.

\subsection{The noise background in the GW data is
  non-Gaussian and nonstationary}
\label{subsec:glitches}
Noise in interferometers arises from a combination of instrumental,
environmental, and anthropomorphic noise sources that are extremely
difficult to characterize 
precisely~\cite{adhikari:14,ligo:09,aasi:15detcharS6,aasi:12detcharVirgo}. Instrumental 
``glitches" can lead to
large excursions over the time-averaged noise and may mimic the
expected time-frequency content of an astrophysical 
signal~\cite{blackburn:08,ligo:09}.  Mitigation strategies against such noise
artifacts include
\begin{enumerate}
\item Coincident observation with multiple,
geographically separated detectors
\item Data quality monitoring and the recording of instrumental 
and environmental vetos derived from auxiliary data channels such as 
seismometers, magnetonometers, etc. 
\item Glitch-detection strategies based on
Bayesian inference (e.g.,~\cite{littenberg:10,powell:15}) or machine learning
(e.g.,~\cite{biswas:13,powell:15}).
\item Using external triggers from EM or neutrino observations to
inform the temporal ``on-source window" in which we expect to find GW
signals and consequently reduce the time period searched.
\end{enumerate}

\subsection{The gravitational wave signal to be expected from a 
  core-collapse event is uncertain} 
The time-frequency characteristics of the GW signal from a core-collapse 
event is strongly dependent on the dominant emission process and
the complex structure, angular momentum distribution, and thermodynamics 
of the progenitor star. In the presence of stochastic emission processes (e.g.,
fluid instabilities such as convection and SASI),
it is impossible to robustly predict the GW signal.
As a result, the optimal method for signal extraction, matched (Wiener)
filtering~\cite{wainstein:62}, cannot be used, as a robust, theoretical 
prediction of the amplitude and phase of the GW signal is required.
Matched filtering is typically used in searches for GWs from compact
binary coalescence, for which robust signal models exist.

The ``excess-power" approach~\cite{anderson:01,vicere:02,sylvestre:03} is 
an alternative to matched filtering for signals of uncertain morphology.
Searching for statistically significant excesses of power in detector 
data in the time-frequency plane, prior information on the sky position, time of 
arrival, and polarization of the targeted GW source can be exploited to reduce 
the noise background and, consequently, the detection false alarm rate.
It can be shown that, in the absence of any knowledge of the signal other than 
its duration and frequency bandwidth, the excess-power method is Neyman-Pearson
optimal in the context of Gaussian noise~\cite{anderson:01}.

\section{Observational Scenarios}
\label{sec:obs_scenarios}
Core-collapse events are the canonical example of multimessenger astrophysical
sources and, as such, are particularly suited to externally triggered GW
searches. In this section, we describe four potential observational scenarios
for CCSNe in the local Universe. 

\subsection{Location of SNe}
We consider CCSNe in four galaxies that contribute 
significantly to the CCSN rate in the Local Group and Virgo cluster. 

The Milky Way, a barred spiral galaxy, is the galaxy that houses our solar
system.  For the purposes of this study, we consider a CCSN in the direction of
the galactic center, at right ascension (RA)
$17^{\mathrm{h}}47^{\mathrm{m}}21.5^{\mathrm{s}}$ and declination 
(Dec) $-5^{\circ}32^{'}9.6^{"}$~\cite{gal2eq}, 
located $\sim9\,\mathrm{kpc}$ from Earth. This is motivated by the work of
Adams~\emph{et al.}~\cite{adams:13}, in which the probability
distribution for the distance of galactic CCSN from Earth is shown to peak around
$\sim9\,\mathrm{kpc}$, and the CCSN location distribution is assumed to trace
the disk of the galaxy. The galactic CCSN rate is estimated
at (0.6-10.5)$\times10^{-2}\,\mathrm{CCSNe}\,\mathrm{yr}^{-1}$~\cite{adams:13}, 
and the youngest known galactic CCSN remnant, Cassiopeia A, is believed to be
$\sim330\,\mathrm{yrs}$ old~\cite{thorstensen:01}. 

The Large Magellanic Cloud (LMC) is home to the most active star-formation
region in the Local Group, the Tarantula Nebula~\cite{kennicutt:86}. Located at RA 
$5^{\mathrm{h}}23^{\mathrm{m}}34.5^{\mathrm{s}}$ and
Dec $-69^{\circ}45^{'}22^{"}$~\cite{devaucoulers:76}, 
the LMC is an irregular galaxy located
$\sim50\,\mathrm{kpc}$ from Earth~\cite{schmidt:92,pietrzynsi:13}, and is 
estimated to have a CCSN rate of
(1.5-3.1)$\times10^{-3}\,\mathrm{CCSNe}\,\mathrm{yr}^{-1}$~\cite{vandenbergh:91,tammann:94}.  
The last CCSN observed in the LMC was SN1987A, a type II-pec SN first detected on 
February 23, 1987, by Kamiokande II via its neutrino burst~\cite{hirata:87}.

The M31 galaxy, also referred to as Andromeda, is the most luminous galaxy in
the Local Group.  Located at RA $0^{\mathrm{h}}42^{\mathrm{m}}44.4^{\mathrm{s}}$
and Dec $41^{\circ}16^{'}8.6^{"}$~\cite{evans:10}, M31 is a spiral galaxy 
located $\sim0.77\,\mathrm{Mpc}$ from 
Earth~\cite{karachentsev:04}, and is estimated to have a CCSN rate of
$\sim2.1\times10^{-3}\,\mathrm{CCSNe}\,\mathrm{yr}^{-1}$~\cite{vandenbergh:91,tammann:94}.  
No CCSNe have yet been observed from M31.

The M82 galaxy, five times brighter than the Milky Way, exhibits starburst
behavior incited by gravitational interaction with M81, a neighboring 
galaxy~\cite{arbutina:07}.  Located at RA
$9^{\mathrm{h}}55^{\mathrm{m}}52.7^{\mathrm{s}}$ and 
Dec $69^{\circ}40^{'}46^{"}$~\cite{jackson:07}, M82 is an 
irregular starburst galaxy at a distance
$\sim3.52\,\mathrm{Mpc}$ from Earth~\cite{jacobs:09}. Its CCSN rate is estimated
to be
$\sim$\,(2.1-20)$\times10^{-2}\,\mathrm{CCSNe}\,\mathrm{yr}^{-1}$~\cite{mattila:01,colina:92}.
The most recent CCSN in M82 was SN2008iz, a Type II SN first observed on May
3, 2008~\cite{brunthaler:10}.

We summarize the relevant information on the aforementioned galaxies in 
Table~\ref{tab:galaxies}.

\begin{table*}
\centering
\begin{tabular}{c|c|c|c|c|c}
Galaxy name & \multicolumn{1}{c}{Right Ascension [Degrees]} &
\multicolumn{1}{c}{Declination [Degrees]} & \multicolumn{1}{c}{Distance [Mpc]}&
CCSN rate [$\times10^{-2}\mathrm{yr}^{-1}$] &
\multicolumn{1}{c}{References} \\
\hline
\hline
Milky Way & 266.42 & -29.01 & 0.01 & $0.6-10.5$ & \cite{adams:13}\\
LMC & \phantom{0}80.89 & -69.76 & 0.05 & $0.1-0.3$ &
\cite{vandenbergh:91,tammann:94,pietrzynsi:13,devaucoulers:76}\\
M31 & \phantom{0}10.69 & \phantom{0}41.27 & 0.77 & $0.2$ &
\cite{vandenbergh:91,tammann:94,evans:10,karachentsev:04}\\
M82 & 148.97 & \phantom{0}69.68 & 3.52 & $2.1-20$ &
\cite{mattila:01,colina:92,jacobs:09,jackson:07}\\
\hline
\hline
\end{tabular}
\caption{Summary of the location, distance, and CCSN rate of the four host 
galaxies considered.}
\label{tab:galaxies}
\end{table*}

\subsection{Analysis times}
\label{subsec:onsourceregions}
The SuperNova Early Warning System
(SNEWS)~\cite{antonioli:04} Collaboration aims to provide a rapid alert 
for a nearby CCSN to the astronomical community, as triggered by neutrino
observations. CCSNe within $\sim100\,\mathrm{kpc}$ will have an
associated neutrino detection. The Large Volume Detector (LVD), a kiloton-scale
liquid scintillator experiment~\cite{aglietta:92}, and Super-Kamiokande
(Super-K), a water-imaging Cerenkov-detector~\cite{ikeda:07} will be able 
to detect neutrinos from a CCSN
with full detection probability ($100\%$) out to $30\,\mathrm{kpc}$ and
$100\,\mathrm{kpc}$, respectively~\cite{agafonova:08,ikeda:07}.
BOREXINO (a 300-ton liquid scintillator experiment~\cite{borexino:09}) is able
to detect all galactic CCSNe~\cite{scholberg:12}, while IceCube 
(a gigaton-scale long string particle
detector made of Antarctic ice~\cite{icecube:11}) can detect a CCSN
in the Large Magellanic Cloud at $6\sigma$ confidence. For 
CCSNe within $\sim0.66\,\mathrm{kpc}$, KamLAND 
(a kiloton-scale liquid scintillator detector~\cite{KamLAND}) will be able 
to detect neutrinos from pre-SN stars at $3\sigma$ confidence~\cite{asakura:15}.

Pagliaroli~\emph{et al.}~\cite{pagliaroli:09} were the first to make
quantitative statements on the use of neutrino detection from CCSNe as external
triggers for an associated GW search, in the context of an analytical
approximation for the anti-electron neutrino luminosity, 
$L_{\bar{\nu}_{e}}$, as a function of 
time. More realistic models for $L_{\nu}$ (see,
e.g.~\cite{fischer:10,mirizzi:15}) suggest that over $\sim95\%$ of the total 
energy in neutrinos is emitted within $\sim10\,\mathrm{s}$ of core bounce. Given 
the neutrino observation time, $t_{0}$, we consider a $60\,\mathrm{s}$ on-source 
window, aligned [$-10,50$]~s about $t_{0}$. We note that a more detailed 
neutrino light curve will allow the time of core bounce
to be localized to $\sim$ few ms~\cite{wallace:15}. This would permit the use 
of a much shorter on-source window, resulting in a lower background rate and
higher detection sensitivity.

For more distant CCSNe, the neutrino burst from core collapse will likely
not be detected, but an EM counterpart will be observed. 
The on-source window derived 
from the EM observation time is dependent on progenitor star characteristics 
(i.e.~progenitor star radius, shock velocity), as well as the observation
cadence. The first EM signature of a CCSN comes at the time of shock
breakout, $t_{\mathrm{SB}}$, when the shock breaks through the stellar
envelope. 

Type Ib and type Ic SNe, hereafter referred to jointly as type Ibc SNe, have very
compact progenitors ($R_{*} \sim $ few (1 - 10)$\Rsun$), and have been
stripped of their stellar envelopes through either intense stellar winds (i.e. Wolf-Rayet
stars), or mass transfer to a binary companion~\cite{filippenko:97,smartt:09}.
Li~\cite{li:07} studied the properties of shock breakout for a variety of type
Ibc SN progenitor models in the context of semianalytic density profiles and 
found shock breakout times in the range
$t_{\mathrm{SB}} \in [1,35]\,\mathrm{s}$. As a conservative estimate, we choose
$t_{\mathrm{SB,min}} = 60\,\mathrm{s}$.

For type II SNe, however, the progenitors are supergiant stars.
Type II-pec SNe, such as SN1987A, have blue supergiant progenitors, with typical
stellar radii of $\sim25\,\Rsun$. More typically, the progenitors are red supergiant 
stars, with typical stellar radii of 
$\sim$\,(100-1000)\,$\Rsun$~\cite{filippenko:97,smartt:09}.
Hydrodynamic simulations of type II-P SN progenitors from 
Bersten \emph{et al.}~\cite{bersten:11} and 
Morozova \emph{et al.}~\cite{morozova:15} show typical breakout times of 
$t_{\mathrm{SB}} \sim$ few $10\,\mathrm{h}$. As a conservative estimate, we
consider the unstripped Type II-P progenitor from Morozova~\emph{et
al.}~\cite{morozova:15}, and use $t_{\mathrm{SB},\mathrm{max}} =
50\,\mathrm{h}$.

In addition to theoretical predictions of the time to shock breakout, the
cadence of observations of the CCSN host galaxy must be considered when deriving
the on-source window. For actively observed galaxies, we expect to have no
greater than $\sim24\,\mathrm{h}$ latency between pre- and post-CCSN observations.
We consider two observational scenarios in which the time scale between pre- and 
post-CCSN images are $t_{\mathrm{obs}} \sim 1\,\mathrm{h}$ and $24\,\mathrm{h}$,
for sources in M31 and M82, respectively. We construct
the on-source window assuming that shock breakout occurs immediately after the
last pre-SN image. Given the time of the last pre-SN observation, the EM trigger
time $t_{0}$, we consider an on-source window of length
$t_{\mathrm{SB}}+t_{\mathrm{obs}}$, aligned 
$[-t_{\mathrm{SB}},t_{\mathrm{obs}}]$ about $t_{0}$.

We summarize the on-source windows used for all observational scenarios
considered in Table~\ref{tab:onsourceregions}.

\begin{table*}
\centering
\begin{tabular}{c|c|c|c}
Galaxy name & Observational counterpart & On-source window for type Ibc [s]&
On-source window for type II [s]\\
\hline
\hline
Milky Way & Neutrino, EM & [-10,+50] & [-10,+50]\\
LMC & Neutrino, EM & [-10,+50] & [-10,+50] \\
M31 & EM & [-60,+3600] & [-180000,+3600] \\
M82 & EM & [-60,+86400] & [-180000,+86400] \\
\hline
\hline
\end{tabular}
\caption{Summary of the observational counterpart used to derive the on-source
window, in addition to the associated on-source window, for type Ibc and type II SNe 
in the four considered host galaxies.}
\label{tab:onsourceregions}
\end{table*}

\begin{figure}[!ht]
\includegraphics[width=\columnwidth]{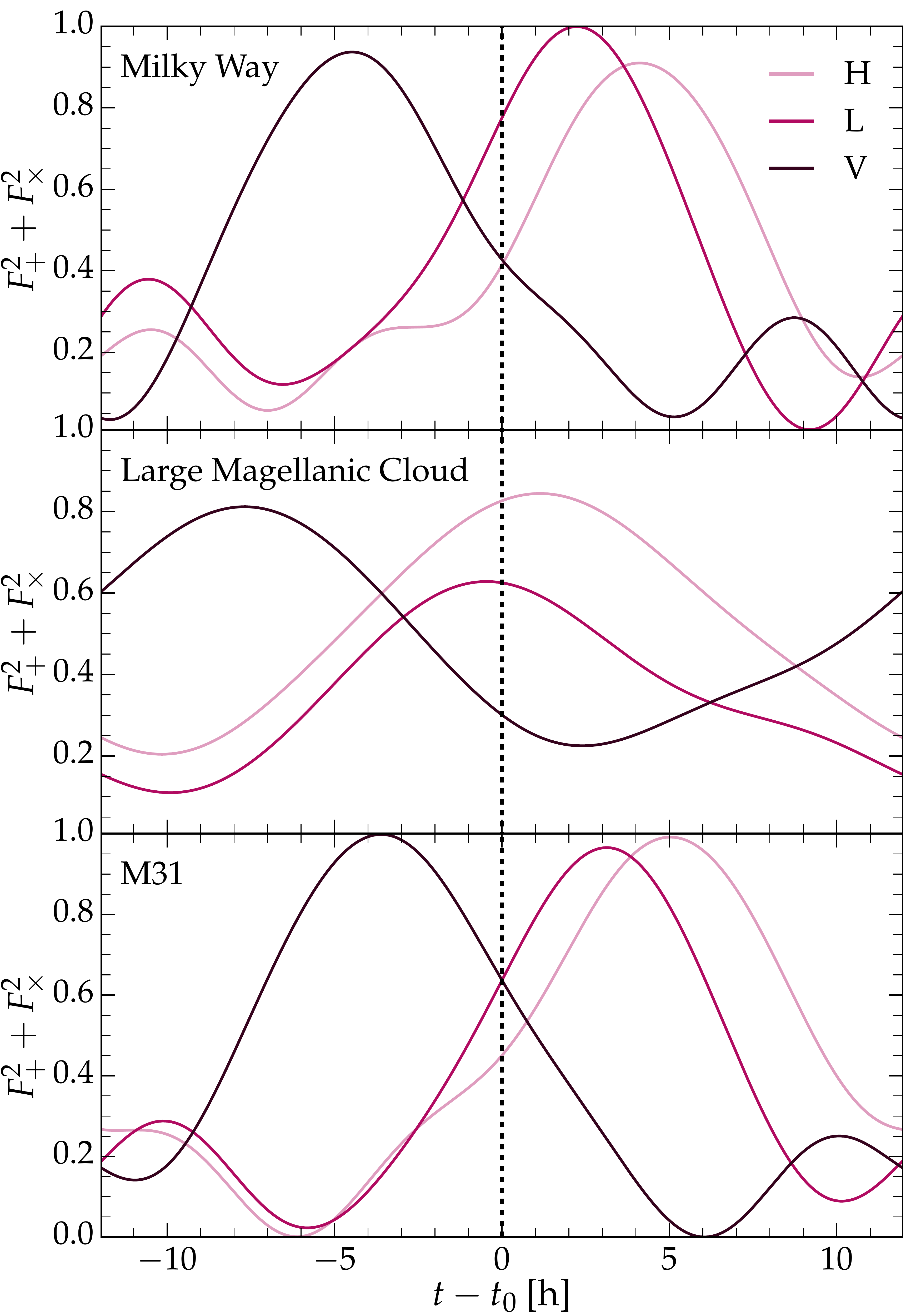}
\caption{The sum-squared antenna response, $F^{2} = F_{+}^{2} +
F_{\times}^{2}$, over one mean sidereal day for the two Advanced LIGO detectors
(H,L), and the Advanced Virgo detector, V, for sources located toward the
Galactic center (top), LMC (middle), and M31 (bottom).
For each galaxy, we indicate the chosen GPS trigger time $t_{0}$
with a dashed black line.}
\label{fig:MW-LMC_antenna}
\end{figure}

The strain detected by a GW interferometer, $h(t)$, is given by
\begin{align}
h(t) = F_{+}(\theta,\Phi,\psi)h_{+}(t) +
F_{\times}(\theta,\Phi,\psi)h_{\times}(t)\,,
\end{align}
where $F_{+,\times}(\theta,\Phi,\psi)$ are the antenna response functions of the 
detector to the two GW polarizations, $h_{+,\times}(t)$. For a
source located at sky position $(\theta,\Phi)$ in detector-centered coordinates,
and characterized by polarization angle $\psi$, $F_{+,\times}$ are given by
\begin{align}
F_{+} = \frac{1}{2}(1+\cos^{2}\theta)\cos2\phi\cos2\psi -
         \cos\theta\sin2\phi\sin2\psi\,,\notag\\
F_{\times} = \frac{1}{2}(1+\cos^{2}\theta)\cos2\phi\sin2\psi -
         \cos\theta\sin2\phi\cos2\psi\,.
\end{align}
The antenna response of the detectors is periodic with an associated
time scale of one sidereal day, due to the rotation of the Earth. As a
consequence, the sensitivity of GW searches using
on-source windows much shorter than this time scale will be strongly dependent on
the antenna response of the detectors to the source location at the relevant
GPS time. In Fig.~\ref{fig:MW-LMC_antenna}, we show the sum-squared antenna response 
for each detector over one sidereal
day, for sources located at the Galactic center, LMC, and M31.
As the sensitivity of the detector network is a function of time, we wish to
choose a central trigger time $t_{0}$ for which the antenna sensitivity is
representative of the average over time. To represent the time-averaged 
sensitivity of the detector network, we choose GPS trigger times
of $t_{0} = 871645255$, $t_{0} = 871784200$, and $t_{0} = 871623913$ for the
Galactic, LMC, and M31 sources, respectively. For CCSNe in M82, 
relying on low-cadence EM triggers, the shortest considered on-source
window is longer than one sidereal day and, as such, the entire range
of antenna responses is encompassed during the on-source window. We 
choose GPS trigger time $t_{0} = 871639563$ for the M82 source, such that
the $74\,\mathrm{h}$ on-source window is covered by the $100\,\mathrm{h}$
stretch of S5 data recolored for this study.

\subsection{Detector networks}
\label{subsec:detectornoise}
As mentioned previously, the GW detector noise will be non-Gaussian and
nonstationary. To this end, we use real GW data from the fifth LIGO science run
(S5) and the first Virgo science run (VSR1), recolored to the target noise
amplitude spectra densities (ASDs)\footnote{The
one-sided amplitude spectral density is the square root of the
one-sided power spectral density,
$S_{h}(f)$.}. for the considered observational scenarios. See 
Sec.~\ref{subsec:recolor} for technical details on the recoloring 
procedure used. 

\begin{figure}[!ht]
\begin{center}
\includegraphics[width=\columnwidth]{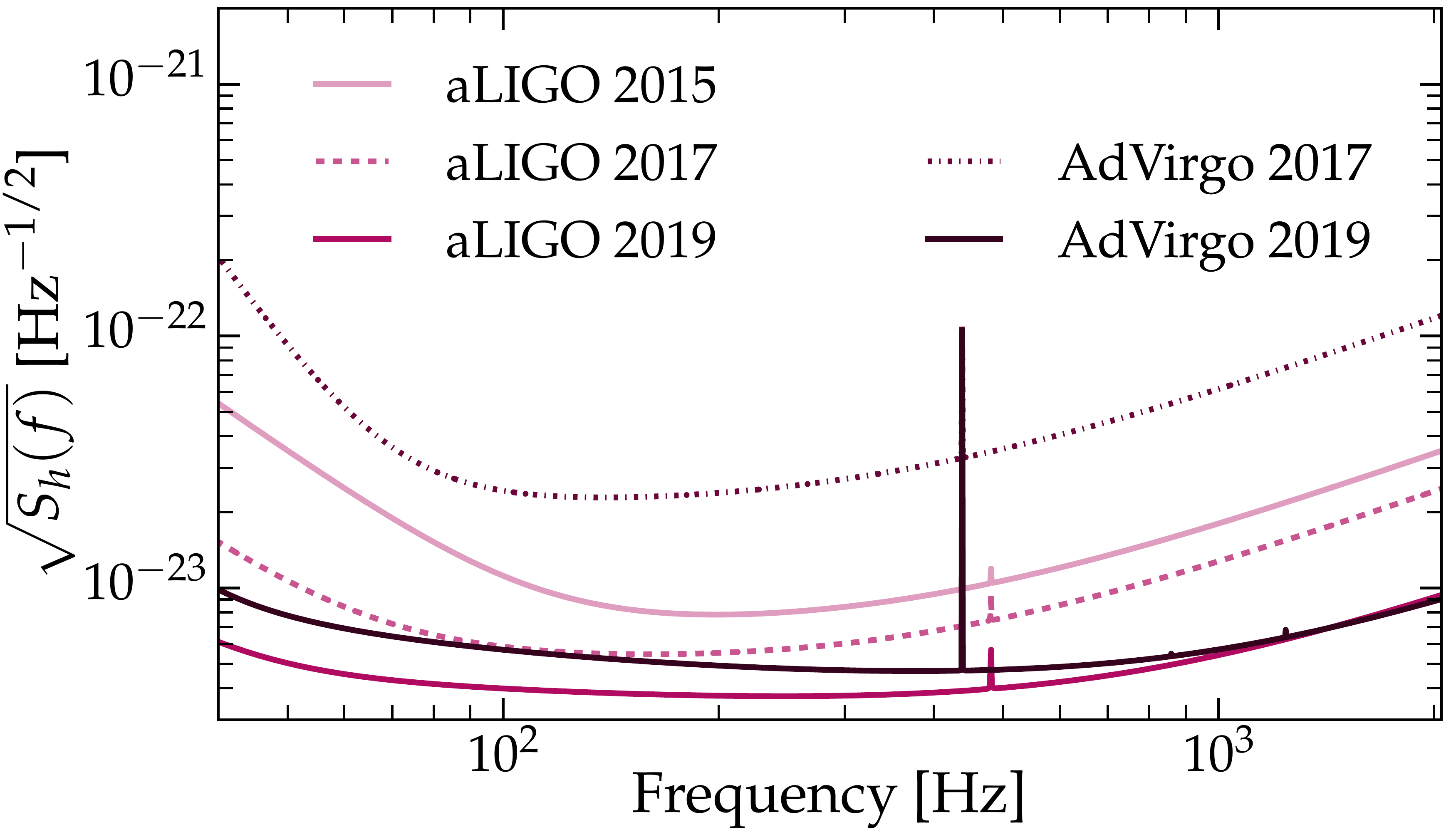}
\end{center}
\caption{The predicted amplitude spectral densities (ASDs), $\sqrt{S_{h}(f)}$ of 
the Advanced LIGO and Advanced Virgo detector noise
for the considered 2015, 2017, and 2019 detector networks~\cite{aasi:13,advirgo}.}
\label{fig:ASDs}  
\end{figure}

We consider a subset of the observing scenarios outlined in Aasi
\emph{et al.}~\cite{aasi:13} to explore how the sensitivity of the 
Advanced detectors to CCSNe will evolve between 2015 and 2019. For all these
cases, we characterize the detector sensitivity by the single-detector binary
neutron star (BNS) range, $d_{\mathrm{R}}$. The BNS range is the standard
figure of merit for detector performance, and is defined as the sky
location- and orientation-averaged distance at which a $(1.4,1.4)\,\Msun$ BNS
system can be detected with an SNR, $\rho \geq 8$. The 2015 scenario assumes a 
two-detector network comprised of the two Advanced 
LIGO detectors (H,L) operating with BNS range $d_{\mathrm{R;HL}} =
54\,\mathrm{Mpc}$ and is hereafter referred to as the \texttt{HL 2015} scenario. 
The 2017 scenario assumes a three-detector network comprised of the two
Advanced LIGO detectors (H,L) operating with BNS range $d_{\mathrm{R;HL}} =
108\,\mathrm{Mpc}$, and the Advanced Virgo detector operating with BNS range of
$d_{\mathrm{R;V}} = 36\,\mathrm{Mpc}$, and is hereafter referred to as the
\texttt{HLV 2017} scenario. In 2019, we consider a three-detector 
network, HLV, with the two Advanced LIGO detectors operating with BNS 
range $d_{\mathrm{R;HL}} = 199\,\mathrm{Mpc}$, and the Advanced Virgo 
detector operating with BNS range $d_{\mathrm{R;V}} = 154\,\mathrm{Mpc}$,
referred to as the \texttt{HLV 2019} observational
scenario~\cite{aasi:13,advirgo}. Figure~\ref{fig:ASDs} 
shows the one-sided ASDs $\sqrt{S_{h}(f)}$ of 
Advanced LIGO and Advanced Virgo as used to recolor the
data for each observational scenario considered.

\section{Gravitational Waves from 
  Core-Collapse Supernovae: Considered Emission Models} 
\label{sec:models}

A broad range of multidimensional processes may emit GWs
during core collapse and the subsequent postbounce CCSN
evolution. These include, but are not necessarily limited to,
turbulent convection driven by negative entropy or lepton gradients
and the SASI (e.g.,
\cite{mueller:13gw,murphy:09,yakunin:10,mueller:e12,marek:09b}),
rapidly rotating collapse and bounce (e.g.,
\cite{dimmelmeier:08,takiwaki:11,ott:12a}), postbounce nonaxisymmetric
rotational instabilities
(e.g.,~\cite{ott:07prl,kuroda:14,scheidegger:10b,corsi:09}), 
rotating collapse to a black hole (e.g.,~\cite{ott:11a}), asymmetric 
neutrino emission and outflows~\cite{yakunin:10,murphy:09,mueller:13gw}, and, 
potentially, rather extreme fragmentation-type instabilities occuring in
accretion torii around nascent neutron stars or black holes~\cite{piro:07}.  A
more extensive discussion of GW emission from CCSNe
can be found in recent reviews on the subject in
Refs.~\cite{ott:09,fryernew:11,kotake:13review}. Most of these emission mechanisms 
source GWs in the most sensitive frequency band of ground-based laser
interferometers ($\sim$\,50-1000\,Hz). Exceptions (and not
considered in this study) are black hole formation
($f_{\mathrm{peak}} \sim\,\mathrm{few}\,\mathrm{kHz}$),
asymmetric neutrino emission, and asymmetric outflows
($f_{\mathrm{peak}} \lesssim 10\,\mathrm{Hz}$).

For the purpose of this study, we consider a subset of the
above GW emission mechanisms and draw example
waveforms from two-dimensional (2D) and three-dimensional (3D) CCSN 
simulations (we refer to these waveforms as numerical waveforms in the
following). In addition, we construct analytical 
phenomenological waveforms that permit us to constrain extreme emission 
scenarios. We consider GW emission in the quadrupole approximation, which has
been shown to be accurate to within numerical error and physical uncertainties
for CCSNe~\cite{reisswig:11ccwave}. In Tables~\ref{tab:numwaveforms} and
\ref{tab:phenomwaveforms}, we summarize key properties of the selected 
numerical and phenomenological 
waveforms, respectively, including the total energy emitted in GWs, $\egw$, the
angle-averaged root-sum-squared GW strain, $\langle h_{\mathrm{rss}}\rangle$,
and the peak frequency of GW emission, $f_{\mathrm{peak}}$. We define $\fpeak$ as
the frequency at which the spectral GW energy density, $\dedflong$, peaks.

We compute $\egw$ as in~\cite{mtw} from the spectral GW energy density,
$\dedflong$, as
\begin{align}
\egw &= \int_{0}^{\infty}\mathrm{d}f\,\dedf\,,
\end{align}
where 
\begin{align}
\dedf &= \frac{2}{5}\frac{G}{c^{5}}\left(2\pi f\right)^{2}\left|\ibarddtilde\right|^{2}\,,
\end{align}
and  
\begin{align}
\ibarddtilde(f) &=
\int_{-\infty}^{\infty}\mathrm{d}t\,\ibardd(t)\,e^{-2\pi ift}\,,
\end{align}
is the Fourier transform of $\ibardd(t)$, the second time derivative of the
mass-quadrupole tensor in the transverse-traceless gauge.
 
To construct the strain for different internal source orientations, we 
present the projection of GW modes, $\hlm$, onto the -$2$ spin-weighted 
spherical harmonic basis, $\sylm$~\cite{ajith:07}. Using this, we may write 
\begin{align}
\hp - i\hc &= \frac{1}{D}\sum_{l=2}^{\infty}\sum_{m=-l}^{l}\hlm\sylm\,,
\label{eq:mode_expansion}
\end{align}
where $(\iota,\phi)$ are the internal source angles describing
orientation.

It has been shown that for CCSN systems, the quadrupole approximation method of
extracting GWs is sufficiently accurate~\cite{reisswig:11ccwave}. As such, we 
consider only the $l = 2$ mode, and can write the mode expansion as
\begin{align}
\htz &= \sqrt{\frac{32\pi}{15}}\frac{G}{c^{4}}\left(\ibarddzz -
\frac{1}{2}\left(\ibarddxx + \ibarddyy \right) \right)\,,\notag\\
\hto &= \sqrt{\frac{16\pi}{5}}\frac{G}{c^{4}}\left(\mp \ibarddxz + i\ibarddyz
\right)\,,\notag\\
\htt &= \sqrt{\frac{4\pi}{5}}\frac{G}{c^{4}}\left(\ibarddxx - \ibarddyy \mp
2i\ibarddxy \right)\,,
\label{eq:quad}
\end{align}
and
\begin{align}
\ytz &= \sqrt{\frac{15}{32\pi}}\sin^{2}\iota\,,\notag\\
\yto &= \sqrt{\frac{5}{16\pi}}\sin\iota \left(1 \pm \cos\iota \right)e^{\pm
i\phi}\,,\notag\\
\ytt &= \sqrt{\frac{5}{64\pi}}\left(1 \pm \cos\iota \right)^{2}e^{\pm 2i\phi}\,.
\end{align}

The root-sum-square strain, $\hrss$, is defined as
\begin{align}
\hrss &= \left[\int_{-\infty}^{\infty}\mathrm{d}t
\left[\hp^{2}(t;\iota,\phi)+\hc^{2}(t;\iota,\phi)\right] \right]^{1/2}\,.
\end{align}
Using the mode decomposition introduced previously, we construct an 
explicit angle-dependent expression for $\hrss$, which we analytically 
average over all source angles.
Defining
\begin{align}
\langle \hrss \rangle &= \iint \mathrm{d}\Omega\, \hrss\,,
\end{align}
we obtain
\begin{align}
\langle \hrss \rangle 
=
\frac{G}{c^{4}}\frac{1}{D}\left[\frac{8}{15}\int_{-\infty}^{\infty}\mathrm{d}t
\left[\ibarddxx^{2} + \ibarddyy^{2} + \ibarddzz^{2} - \right.\right.\notag\\
\left.\left.(\ibarddxx\ibarddyy + \ibarddxx\ibarddzz + \ibarddyy\ibarddzz) 
+ 3\left(\ibarddxy^{2} + \ibarddxz^{2} + \ibarddyz^{2}\right) \right] \right]^{1/2}\,.
\label{eq:hrss}
\end{align}

\subsection{Numerical waveforms}
\label{subsec:numwaveforms}
\subsubsection{Gravitational waves from convection and SASI}
\label{subsubsec:nuCC}
Postbounce CCSN cores are unstable to convection. The stalling shock
leaves behind an unstable negative entropy gradient, leading to
a burst of prompt convection soon after core bounce. As the
post-bounce evolution
proceeds, neutrino heating sets up a negative entropy gradient in the
region of net energy deposition (the gain layer) behind the
shock, leading to neutrino-driven convection. Simultaneously,
neutrino diffusion establishes a negative lepton gradient in the
mantle of the proto-neutron star (NS), leading to proto-NS 
convection. The GW signal from these convective
processes has a broad spectrum. The prompt convection
GW emission occurs at frequencies in the
range $100-300\,\mathrm{Hz}$, while neutrino-driven convection at later times 
sources GW emission with significant power at
frequencies between $\sim$ 300-1000\,Hz (increasing with
time~\cite{murphy:09,mueller:13gw,yakunin:10,yakunin:15}). Proto-NS 
convection contributes at the highest frequencies 
($\gtrsim 1000\,\mathrm{Hz}$). While the frequency 
content of the signal is robust, the phase is stochastic
due to the chaotic nature of turbulence~\cite{ott:09,kotake:09}. 

In addition to convection, depending on
progenitor structure (and, potentially, dimensionality of the
simulation; cf.~\cite{burrows:12,ott:13a,mueller:12b,hanke:13,couch:15}), the
shock front may become unstable to SASI, which leads to large-scale
modulations of the accretion flow. This results in sporadic large
amplitude spikes in the GW signal when large
accreting plumes are decelerated at the edge of the proto-NS
(e.g.,~\cite{murphy:09,mueller:13gw}).

\begin{figure}
\includegraphics[width=\columnwidth]{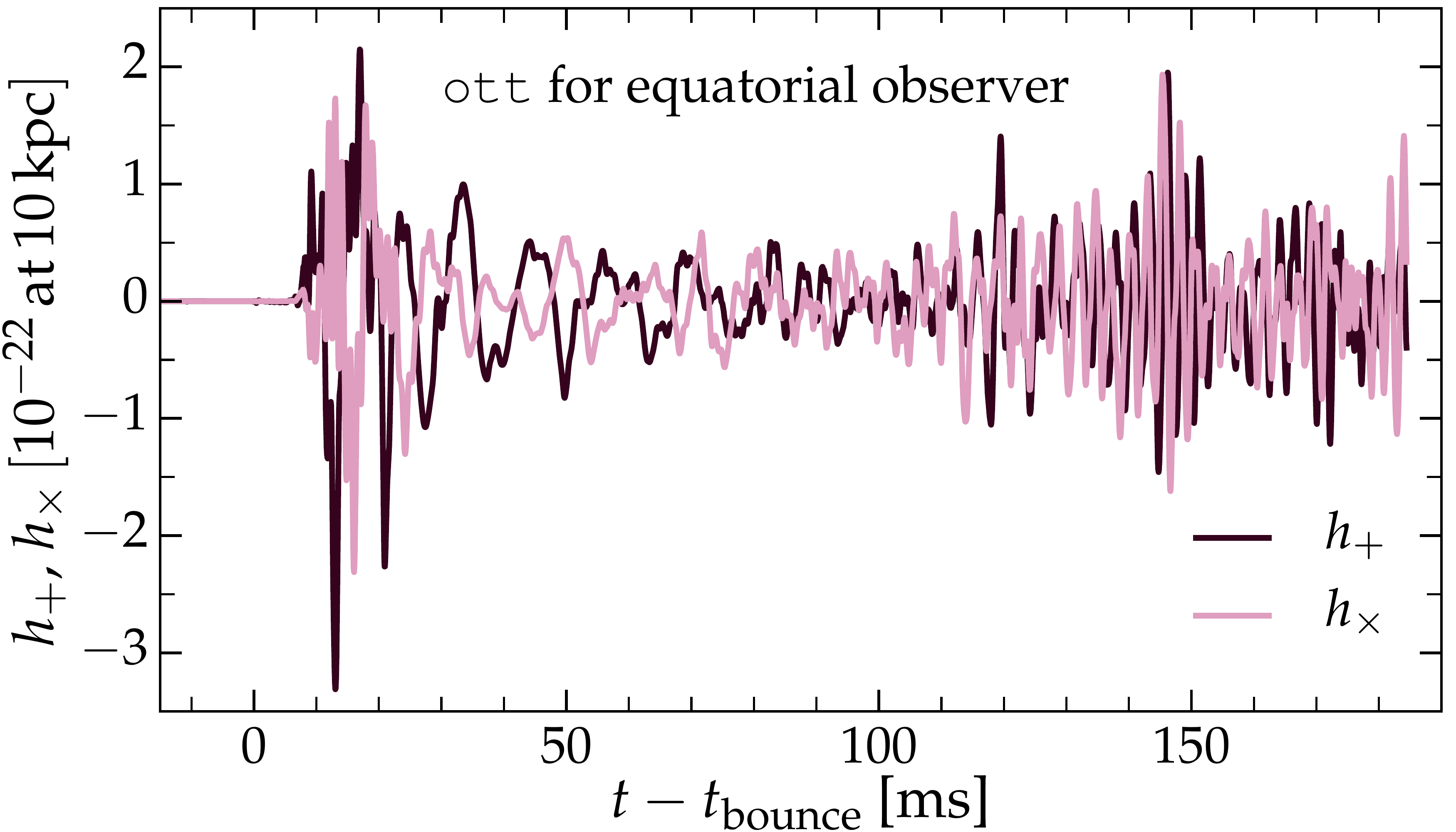}
\includegraphics[width=\columnwidth]{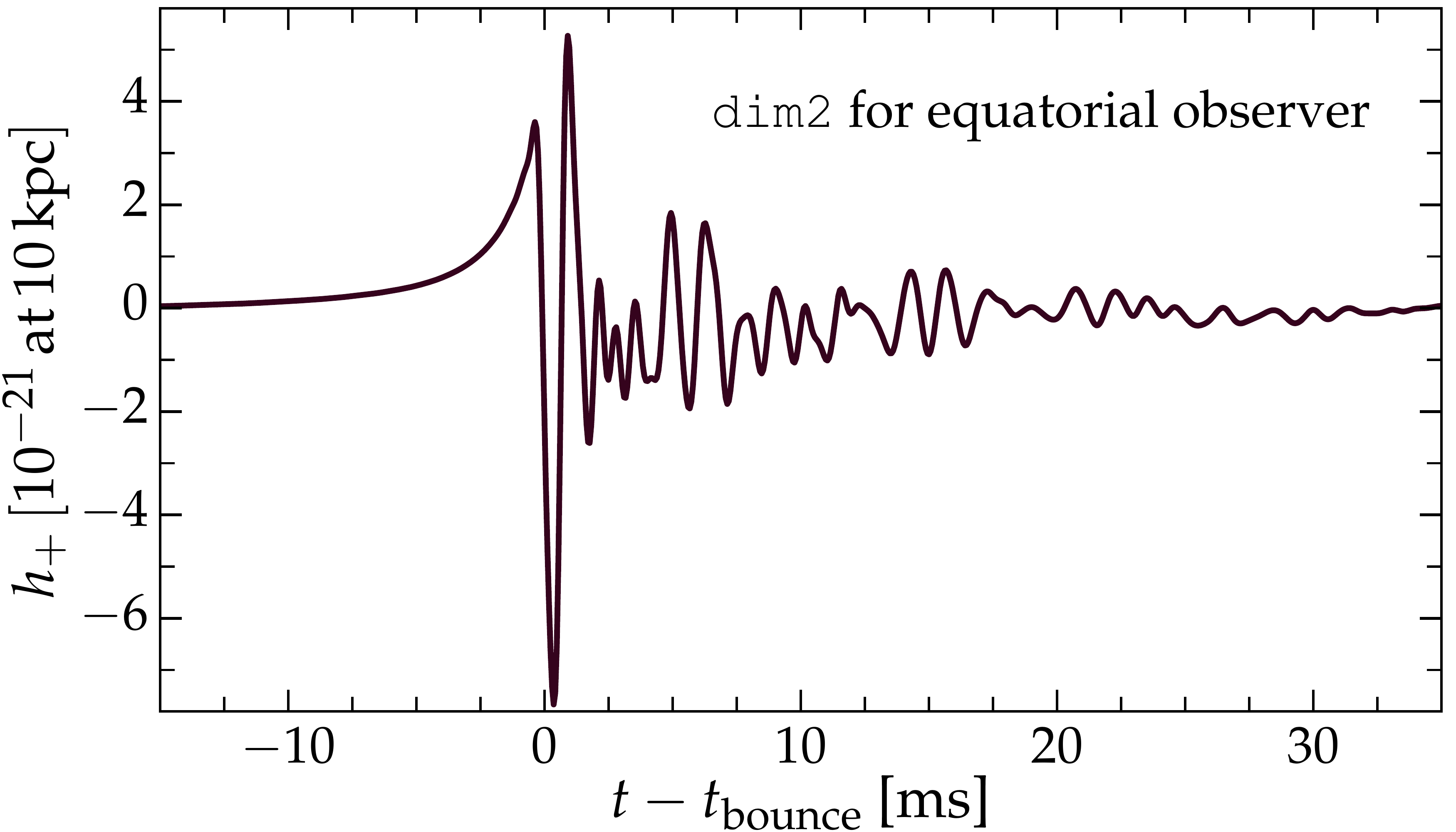}
\includegraphics[width=\columnwidth]{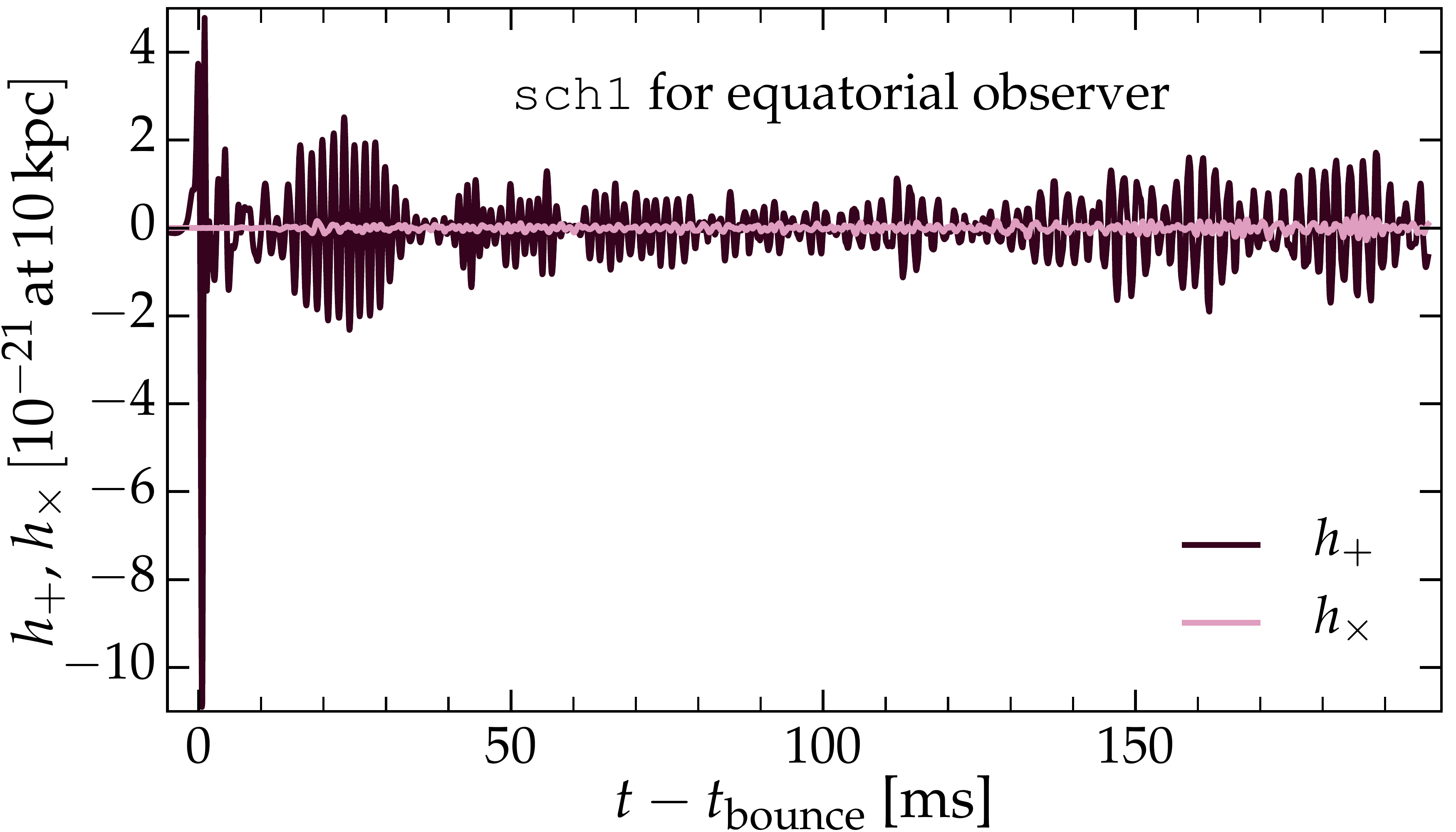}
\caption{The time domain GW strain for representative models of convection and
standing accretion-shock instability (\texttt{ott}; top panel), bounce and 
ringdown of the proto-neutron star (\texttt{dim2}; middle panel), and 
nonaxisymmetric rotational instabilities (\texttt{sch1}; bottom panel) as 
seen by an equatorial ($\iota=\pi/2$; $\phi=0$)
observer at $10\,\mathrm{kpc}$. We note that the typical GW strain from 
rotating core collapse is roughly an order of magnitude larger than the 
typical GW strain from neutrino-driven explosions. In addition, the typical 
GW signal duration of bounce and ringdown of the proto-neutron star is 
$\sim$\,few 10\,ms, compared to the typical GW signal duration of 
$\sim$\,few 100\,ms for neutrino-driven explosions. Nonaxisymmetric 
rotational instabilities, however, may persist for $\sim$\,few 100\,ms.}
\label{fig:numwaveforms}
\end{figure}

We draw sample waveforms for GWs from nonrotating core collapse 
from the studies of Yakunin~\emph{et al.}~\cite{yakunin:10},
M\"uller~\emph{et al.}~\cite{mueller:e12}, and 
Ott~\emph{et al.}~\cite{ott:13a}. 
Yakunin~\emph{et al.}\ performed 2D simulations of neutrino-driven 
CCSNe. We choose a waveform obtained from the simulation of a $15\,M_\odot$ 
progenitor star (referred to as \texttt{yak} in the following). Due to
axisymmetry, the extracted waveform is linearly polarized.
M\"uller~\emph{et al.} performed 3D simulations of
neutrino-driven CCSNe with a number of approximations to make the
simulations computationally feasible.  Importantly, they started their
simulations after core bounce and assumed a time-varying inner
boundary, cutting out much of the proto-neutron star. Prompt and
proto-neutron star convection do not contribute to their
waveforms, and higher frequency GW emission is suppressed due to the
artificial inner boundary. As the simulations are 3D, the M\"uller~\emph{et al.}\
waveforms have two polarizations, and we use waveforms of 
models L15-3, W15-4 (two different $15\,\Msun$ progenitors), and
N20-2 (a $20\,M_\odot$ progenitor). We refer to these waveforms as
\texttt{m\"uller1}, \texttt{m\"uller3}, and \texttt{m\"uller2}, 
respectively. Ott~\emph{et al.}\ performed 3D
simulations of neutrino-driven CCSNe. The simulations are general-relativistic
and incorporate a three-species neutrino leakage scheme.
As the simulations are 3D, the Ott~\emph{et
al.}\ waveforms have two polarizations, and we use the GW waveform from
model $s27f_{\mathrm{heat}}1.05$ (a $27\,\Msun$ progenitor). We hereafter 
refer to this waveform as \texttt{ott}. We plot the GW signal for the 
\texttt{ott} model in the top panel of Fig.~\ref{fig:numwaveforms}.

\begin{table*}
\begin{tabular}{lclcll}
\multicolumn{1}{c}{Waveform type}
&\multicolumn{1}{c}{Ref.}
&\multicolumn{1}{c}{Waveform name}
&\multicolumn{1}{c}{$\langle h_\mathrm{rss} \rangle$}
&\multicolumn{1}{c}{$f_\mathrm{peak}$}
&\multicolumn{1}{c}{$\egw$}\\
&&
&\multicolumn{1}{c}{$[10^{-22}$ at $10\,\mathrm{kpc}]$}
&\multicolumn{1}{c}{$[\mathrm{Hz}]$}
&\multicolumn{1}{c}{$[M_{\odot} c^{2}]$}\\
\hline
\hline
2D neutrino-driven convection and SASI
&\cite{yakunin:10}
&\texttt{yak}
&$1.89$
&$\phantom{0}888$
&$9.08\times10^{-9}$\\
3D neutrino-driven convection and SASI
&\cite{mueller:e12}
&\texttt{m\"uller1}
&$1.66$
&$\phantom{0}150$
& $3.74\times10^{-11}$\\
3D neutrino-driven convection and SASI
&\cite{mueller:e12}
&\texttt{m\"uller2}
&$3.85$
&$\phantom{0}176$
& $4.37\times10^{-11}$ \\
3D neutrino-driven convection and SASI
&\cite{mueller:e12}
&\texttt{m\"uller3}
&$1.09$
&$\phantom{0}204$
& $3.25\times10^{-11}$\\
3D neutrino-driven convection and SASI
&\cite{ott:13a}
&\texttt{ott}
&$0.24$
&$1019$
& $7.34\times10^{-10}$\\
\hline
2D rotating core collapse
&\cite{dimmelmeier:08}
&\texttt{dim1}
&$1.05$
&$\phantom{0}774$
&$7.69\times10^{-9}$\\
2D rotating core collapse
&\cite{dimmelmeier:08}
&\texttt{dim2}
&$1.80$
&$\phantom{0}753$
&$2.79\times10^{-8}$\\
2D rotating core collapse
&\cite{dimmelmeier:08}
&\texttt{dim3}
&$2.69$
&$\phantom{0}237$
&$1.38\times10^{-9}$\\
3D rotating core collapse
&\cite{scheidegger:10}
&\texttt{sch1}
&$5.14$
&$\phantom{0}465$
& $2.25\times10^{-7}$\\
3D rotating core collapse
&\cite{scheidegger:10}
&\texttt{sch2}
&$5.80$
&$\phantom{0}700$
& $4.02\times10^{-7}$\\
\hline
\hline
\end{tabular}
\caption{Key characteristics of ``numerical" waveforms from multidimensional
CCSN simulations. $E_{\mathrm{GW}}$ is the energy emitted in GWs, $\langle
h_{\mathrm{rss}} \rangle$ is the angle-averaged root-sum-square strain
[Eq.~(\ref{eq:hrss})], and $f_{\mathrm{peak}}$ is the frequency at which the
spectral GW energy $\dedflong$ peaks.}
\label{tab:numwaveforms}
\end{table*}


\subsubsection{Gravitational waves from rotating core collapse and bounce}
\label{subsubsec:rotCC}
Rotation leads to oblateness (an $\ell = 2, m=0$ quadrupole
deformation) of the inner quasihomologously collapsing core. Extreme
accelerations experienced by the inner core at bounce lead to a large spike in
the GW signal at bounce, followed by ringdown of the proto-neutron star as it
settles to its new equilibrium state (see,
e.g.,~\cite{dimmelmeier:08,ott:09,fuller:15} 
for a detailed discussion). The GW signal is dependent on
the mass of the inner core, its angular momentum distribution, and the equation
of state of nuclear matter. There are significant
uncertainties in these  and it is difficult to exactly predict 
the time series of the GW signal. Nevertheless, work by several
authors~\cite{summerscales:08,heng:09,roever:09,logue:12,abdikamalov:14,engels:14,fuller:15} 
has demonstrated that GW emission from rotating core collapse and
bounce has robust features that can be identified and used to infer properties of
the progenitor core.

We draw three sample waveforms from the axisymmetric
general-relativistic (conformally flat) simulations of Dimmelmeier~\emph{et
  al.}~\cite{dimmelmeier:08}. All were performed with a $15$-$M_\odot$
progenitor star and the Lattimer-Swesty equation of state~\cite{lseos:91}. 
The three linearly polarized waveforms drawn from~\cite{dimmelmeier:08}, 
s15A2O05-ls, s15A2O09-ls, and s15A3O15-ls, differ
primarily by their initial rotation rate and angular momentum
distribution. We refer to them as \texttt{dim1} (slow and rather
uniform precollapse rotation), \texttt{dim2} (moderate and rather
uniform precollapse rotation), and \texttt{dim3} (fast and strongly
differential precollapse rotation), respectively.

Shortly after core bounce, nonaxisymmetric rotational instabilities
driven by rotational shear (e.g.,
\cite{fryer:02,rotinst:05,ott:07prl,scheidegger:10,kuroda:14}) or, in the limit
of extreme rotation, by a classical high-$T/|W|$ instability at $T/|W|
\gtrsim 25-27\%$ \cite{baiotti:07}, where $T$ is the rotational
kinetic energy and $W$ is the gravitational energy, may set in. The
nonaxisymmetric deformations may lead to a signficant enhancement of
the GW signal from the postbounce phase of rotating
CCSNe.  We choose two sample waveforms from the 3D Newtonian,
magnetohydrodynamical simulations of Scheidegger~\emph{et
al.}~\cite{scheidegger:10}, which use a neutrino leakage scheme. All 
were performed with a $15\,\Msun$ progenitor
star, and the Lattimer-Swesty equation of state~\cite{lseos:91}. Due to the 3D
nature of the simulations, the Scheidegger~\emph{et al.} waveforms have 
two polarizations. We employ waveforms for models
$\mathrm{R3E1AC}_{\mathrm{L}}$ (moderate precollapse rotation, toroidal/poloidal
magnetic field strength of $10^{6}\,\mathrm{G}$/$10^{9}\,\mathrm{G}$), and
$\mathrm{R4E1FC}_{\mathrm{L}}$ (rapid precollapse rotation, toroidal/poloidal
magnetic field strength of $10^{12}\,\mathrm{G}$/$10^{9}\,\mathrm{G}$).  We 
hereafter refer to these waveforms as \texttt{sch1} and \texttt{sch2}, respectively.

\subsection{Phenomenological waveforms}
\label{subsec:anawaveforms}
\subsubsection{Gravitational waves from long-lived rotational instabilities}
\label{subsubsec:bars}
Proto-neutron stars with ratio of rotational kinetic energy $T$ to
gravitational energy $|W|$, $\beta = T/|W| \gtrsim$ 25-27\% become dynamically
unstable to nonaxisymmetric deformation (with primarily $m=2$ bar
shape). If $\beta \gtrsim 14\%$, an instability may grow on a secular
(viscous, GW backreaction) time scale, which may be
seconds in proto-neutron stars (e.g.,~\cite{lai:01}). Furthermore,
proto-neutron stars are born differentially rotating (e.g.,
\cite{ott:06spin}) and may thus be subject to a dynamical shear
instability driving nonaxisymmetric deformations that are of smaller
magnitude than in the classical instabilities, but are likely to
set in at much lower $\beta$. Since this instability operates
on differential rotation, it may last for as long as accretion
maintains sufficient differential rotation in the outer proto-neutron star
(e.g.,~\cite{balbinski:85,watts:05,rotinst:05,ott:07prl,scheidegger:10,kuroda:14}
and references therein).

For simplicity, we assume that the net result of all these
instabilities is a bar deformation, whose GW emission
we model in the Newtonian quadrupole approximation for a cylinder of
length $l$, radius $r$ and mass $M$ in the $x$-$y$ plane, rotating about
the $z$ axis. We neglect spin-down via GW backreaction. The second 
time derivative of the bar's reduced mass-quadrupole tensor is given by
\begin{align}
\ibardd &= 
\frac{1}{6} M (l^2 - 3 r^2) \,\Omega^2 \left(\begin{array}{rr}
-\cos{2\Omega t}&\sin{2 \Omega t}\\
\sin{2 \Omega t}&\cos{2 \Omega t}
\end{array}\right)\,\,,
\end{align}
where $\Omega = 2\pi f$ is the angular velocity of the bar 
(see, e.g.,~\cite{ott:10dcc} for details).  We then obtain the GW signal using
the quadrupole formula in Eq.~(\ref{eq:quad})~\cite{thorne:87,ajith:07}.

We generate representative analytic bar waveforms by fixing the bar
length to $60\,\mathrm{km}$, its radius to $10\,\mathrm{km}$ and
varying the mass in the deformation $M$, the spin frequency $f$, and duration of
the bar mode instability $\Delta t$. In practice, we
scale the waveforms with a Gaussian envelope $\propto \exp(- (t-\Delta
t)^2 / (\Delta t / 4)^2$) to obtain nearly zero amplitudes at start
and end of the waveforms, resulting in waveforms of sine-Gaussian
morphology. In this study, we consider three bars of mass $M =
0.2\,\Msun$, with $(f,\Delta t) = (400\,\mathrm{Hz},0.1\,\mathrm{s})$,
$(400\,\mathrm{Hz},1\,\mathrm{s})$, and $(800\,\mathrm{Hz},0.1\,\mathrm{s})$
(hereafter referred to as \texttt{longbar1}, \texttt{longbar2}, and
\texttt{longbar3}, respectively), and three bars of mass $M = 1\,\Msun$ with
$(f,\Delta t) = (400\,\mathrm{Hz},0.1\,\mathrm{s})$,
$(400\,\mathrm{Hz},1\,\mathrm{s})$, and $(800\,\mathrm{Hz},0.025\,\mathrm{s})$
(hereafter referred to as \texttt{longbar4}, \texttt{longbar5}, and
\texttt{longbar6}, respectively). We choose these parameters to explore the 
regime of strong bar-mode GW emission with the constraint
that the strongest signal must emit less energy than is available in
collapse, $E_\mathrm{GW} \lesssim 0.15 M_\odot c^2$. Values of $\langle
h_{\mathrm{rss}}\rangle$, $f_{\mathrm{peak}}$, and $E_{\mathrm{GW}}$ for the
six representative waveforms used in this study are shown in 
Table~\ref{tab:phenomwaveforms}.

\begin{figure}
\includegraphics[width=\columnwidth]{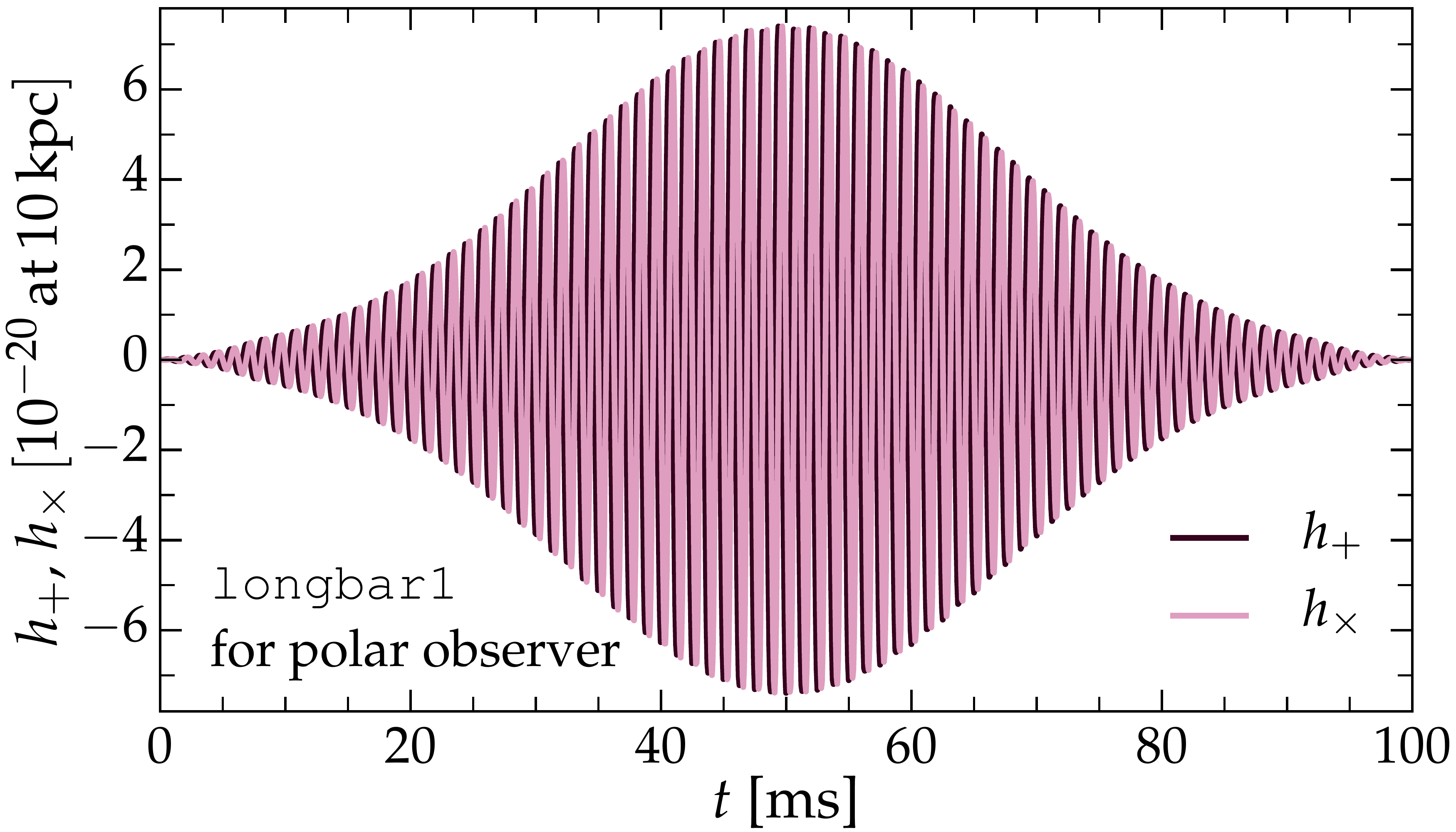}
\includegraphics[width=\columnwidth]{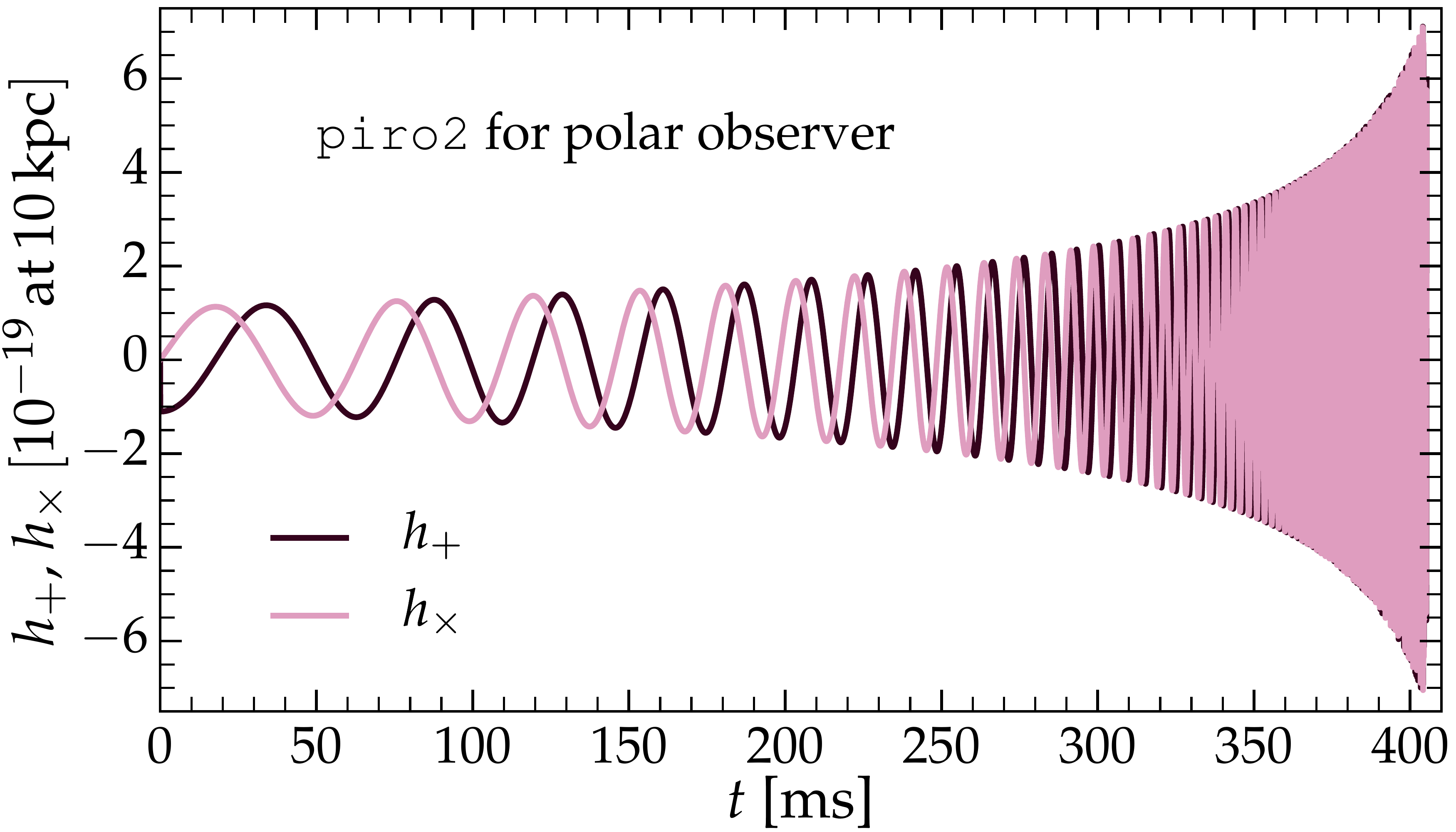}
\caption{The time domain GW strain for
representative models of bar-mode instability (\texttt{longbar1}; top panel) and
disk fragmentation instability (\texttt{piro2}; bottom panel), as seen 
by a polar ($\iota = 0$; $\phi = 0$) observer at $10\,\mathrm{kpc}$.}
\label{fig:phenomwaveforms}
\end{figure}

\begin{table*}
\begin{tabular}{lclcrl}
\multicolumn{1}{c}{Waveform type}
&\multicolumn{1}{c}{Ref.}
&\multicolumn{1}{c}{Waveform name}
&\multicolumn{1}{c}{$\langle h_\mathrm{rss} \rangle$}
&\multicolumn{1}{c}{$f_\mathrm{peak}$}
&\multicolumn{1}{c}{$\egw$}\\
&&
&\multicolumn{1}{c}{$[10^{-20}$ at $10\,\mathrm{kpc}]$}
&\multicolumn{1}{c}{$[\mathrm{Hz}]$}
&\multicolumn{1}{c}{$[M_{\odot} c^{2}]$}\\
\hline
\hline
Long-lasting bar mode
&\cite{ott:10dcc}
&\texttt{longbar1}
&$\phantom{0}1.48$
&\phantom{0}800
&$2.98\times10^{-4}$\\
Long-lasting bar mode
&\cite{ott:10dcc}
&\texttt{longbar2}
&$\phantom{0}4.68$
&\phantom{0}800
&$2.98\times10^{-3}$\\
Long-lasting bar mode
&\cite{ott:10dcc}
&\texttt{longbar3}
&$\phantom{0}5.92$
&1600
&$1.90\times10^{-2}$\\
Long-lasting bar mode
&\cite{ott:10dcc}
&\texttt{longbar4}
&$\phantom{0}7.40$
&\phantom{0}800
&$7.46\times10^{-3}$\\
Long-lasting bar mode
&\cite{ott:10dcc}
&\texttt{longbar5}
&$23.41$
&\phantom{0}800
&$7.45\times10^{-2}$\\
Long-lasting bar mode
&\cite{ott:10dcc}
&\texttt{longbar6}
&$14.78$
&1600
&$1.18\times10^{-1}$\\ 
\hline
Torus fragmentation instability
&\cite{piro:07}
&\texttt{piro1}
&$\phantom{0}2.55$
&2035
&$6.77\times10^{-4}$\\
Torus fragmentation instability
& \cite{piro:07}
&\texttt{piro2}
&$\phantom{0}9.94$
&1987
&$1.03\times10^{-2}$\\
Torus fragmentation instability
& \cite{piro:07}
&\texttt{piro3}
&$\phantom{0}7.21$
&2033
&$4.99\times10^{-3}$\\
Torus fragmentation instability
& \cite{piro:07}
&\texttt{piro4}
&$28.08$
&2041
&$7.45\times10^{-2}$\\
\hline
\hline
\end{tabular}
\caption{Key characteristics of the considered waveforms from phenomenological 
models. $E_{\mathrm{GW}}$ is the energy emitted in GWs,
$\langle h_{\mathrm{rss}} \rangle$ is the angle-averaged root-sum-square strain
[Eq.~(\ref{eq:hrss})], and $f_{\mathrm{peak}}$ is the frequency at which the
spectral GW energy density $\dedflong$ peaks.}
\label{tab:phenomwaveforms}
\end{table*}

\subsubsection{Disk fragmentation instability}
\label{subsubsec:piro}
If the CCSN mechanism fails to reenergize the
stalled shock (see, e.g.,~\cite{janka:07}), the proto-neutron star
will collapse to a black hole on a
time scale set by accretion~(e.g.,~\cite{oconnor:11}). 
Provided sufficient angular momentum, a massive
self-gravitating accretion disk/torus may form around the nascent
stellar-mass black hole with mass $M_\mathrm{BH}$. This scenario may
lead to a collapsar-type gamma-ray burst (GRB) or an engine-driven
SN~\cite{wb:06}.

The inner regions of the disk are geometrically thin due to efficient
neutrino cooling, but outer regions are thick and may be
gravitationally unstable to fragmentation at large 
radii~\cite{piro:07,perna:06}.  We follow work by Piro and 
Pfahl~\cite{piro:07}, and consider the case in which a single
gravitationally bound fragment forms in the disk and collapses to a
low-mass neutron star with $M_f \sim 0.1-1 M_\odot \ll
M_\mathrm{BH}$. We then obtain the predicted GW signal using
Eq.~(\ref{eq:quad})~\cite{thorne:87,ajith:07}, assuming the fragment is orbiting
in the ($x$-$y$)-plane, such that
\begin{align}
\ibardd = 2 \frac{M_\mathrm{BH} M_f}{(M_\mathrm{BH} + M_f)} r^2\, \Omega^2 
\left(\begin{array}{rr}
-\cos{2\Omega t}&-\sin{2 \Omega t}\\
-\sin{2 \Omega t}&\cos{2 \Omega t}
\end{array}\right)\,\,.
\end{align}
For more technical details, including the waveform generation code, we direct
the reader to~\cite{piro:07,santamaria:11dcc}. We consider waveforms from four 
example systems with
$(M_{\mathrm{BH}},M_{f}) = (5\,\Msun,0.07\,\Msun)$, 
$(5\,\Msun,0.58\,\Msun)$, $(10\,\Msun,0.14\,\Msun)$,
and $(10\,\Msun,1.15\,\Msun)$ (hereafter denoted \texttt{piro1}, \texttt{piro2},
\texttt{piro3}, and \texttt{piro4}, respectively). Values of $\langle
h_{\mathrm{rss}}\rangle$, $f_{\mathrm{peak}}$, and $E_{\mathrm{GW}}$ for the
four representative waveforms used in this study are shown in 
Table~\ref{tab:phenomwaveforms}.

\subsubsection{Ad hoc signal models}
\label{subsubsec:sg}
It is possible that there are GW emission mechanisms from CCSNe that we have
not considered. In this case, it is instructive to determine the sensitivity of 
our GW search to short, localized bursts of GWs in time-frequency
space. For this reason, we include \emph{ad hoc} signal
models in our signal injections, in addition to the aforementioned physically
motivated signal models. We take motivation from the all-sky, all-time searches for GW
bursts performed in the intial detector
era~\cite{abadie:12allskyS5,abadie:12s6burst}, and consider linearly and
elliptically polarized sine-Gaussian GW bursts. Characterized by central 
frequency, $f_{0}$, and quality factor, $Q$, the strain is given by
\begin{align}
h_{+}(t) &= A\left(\frac{1 +
\alpha^{2}}{2}\right)\exp(-2\pi f_{0}^{2}t^{2}/Q^{2})\sin(2\pi
f_{0}t)\,,\notag \\
h_{\times}(t) &= A\alpha \exp(-2\pi f_{0}^{2}t^{2}/Q^{2})\cos(2\pi f_{0}t)\,,
\label{eq:sg}
\end{align}
where $A$ is some common scale factor, and $\alpha$ is the ellipticity, where
$\alpha = 0$ and $1$ for linearly and circularly polarized waveforms
respectively. Assuming isotropic energy emission, we may compute the
energy in GWs associated with a sine-Gaussian burst as
\begin{align}
E_{\mathrm{GW}} &= \frac{\pi^{2}c^{3}}{G}d^{2}f_{0}^{2}h_{\mathrm{rss}}^{2}\,,
\label{eq:egw_sg}
\end{align}
where $d$ is the distance at which $h_{\mathrm{rss}}$ is computed.
In Table~\ref{tab:sg}, we list the $f_{0}$, $Q$, and $\alpha$
values for all sine-Gaussian waveforms considered in this study.

\begin{table}
\begin{tabular}{lrrl}
\multicolumn{1}{c}{Model Name}
&\multicolumn{1}{c}{$f_{0}$ $[\mathrm{Hz}]$}
&\multicolumn{1}{c}{$Q$}
&\multicolumn{1}{c}{$\alpha$} \\
\hline
\hline
\texttt{sglin1},\texttt{sgel1}
& 70
& 3
& 0,1\\
\texttt{sglin2},\texttt{sgel2}
& 70
& 9
& 0,1\\
\texttt{sglin3},\texttt{sgel3}
& 70
& 100
& 0,1\\
\texttt{sglin4},\texttt{sgel4}
& 100
& 9
& 0,1\\
\texttt{sglin5},\texttt{sgel5}
& 153
& 9
& 0,1\\
\texttt{sglin6},\texttt{sgel6}
& 235
& 3
& 0,1\\
\texttt{sglin7},\texttt{sgel7}
& 235
& 9
& 0,1\\
\texttt{sglin8},\texttt{sgel8}
& 235
& 100
& 0,1\\
\texttt{sglin9},\texttt{sgel9}
& 361
& 9
& 0,1\\
\texttt{sglin10},\texttt{sgel10}
& 554
& 9
& 0,1\\
\texttt{sglin11},\texttt{sgel11}
& 849
& 3
& 0,1\\
\texttt{sglin12},\texttt{sgel12}
& 849
& 9
& 0,1\\
\texttt{sglin13},\texttt{sgel13}
& 849
& 100
& 0,1\\
\texttt{sglin14},\texttt{sgel14}
& 1053
& 9
& 0,1\\
\texttt{sglin15},\texttt{sgel15}
& 1304
& 9
& 0,1\\
\hline
\hline
\end{tabular}
\caption{Key characteristics of the \emph{ad hoc} sine-Gaussian waveforms
employed in this study. $f_{0}$ is the central frequency, $Q$ is 
the quality factor, and $\alpha$ is the ellipticity. See Eq.~(\ref{eq:sg})
in Sec.~\ref{subsubsec:sg} for details.}
\label{tab:sg}
\end{table}


\section{Data Analysis Methods}
\label{sec:GWdetection}

\subsection{X-Pipeline: A search algorithm for gravitational wave bursts}
\label{subsec:x}

\texttt{X-Pipeline} is a coherent analysis pipeline used to search for GW 
transient events associated with CCSNe, gamma-ray bursts (GRBs), and other 
astrophysical triggers. \texttt{X-Pipeline} has a number of features 
designed specifically to address the challenges discussed in
Sec.~\ref{sec:challenges}. For example, since the signal duration is
uncertain, \texttt{X-Pipeline} uses multiresolution Fourier transforms 
to maximize sensitivity across a range of possible signal durations. The 
pixel clustering procedure applied to time-frequency maps of the data is 
designed to find arbitrarily shaped, connected events~\cite{was:11thesis}.
The potentially nonstationary data is whitened in blocks of 256\,s duration,
removing the effect of variations in background noise levels which typically 
happen on longer time scales. Short-duration noise glitches are removed by 
comparing measures of interdetector correlations to a set of thresholds 
that are tuned using simulated GW signals from the known sky position of 
the CCSNe and actual noise glitches over the on-source window. The 
thresholds are selected to satisfy the Neyman-Pearson optimality 
criterion (maximum detection efficiency at fixed false-alarm probability), 
and are automatically adjusted for the event amplitude to give robust 
rejection of loud glitches. We provide a brief overview of the 
functionality of \texttt{X-Pipeline} here, specifically in the context 
of CCSN searches, and direct the reader to the \texttt{X-Pipeline} 
technical document for a more in-depth description~\cite{sutton:10}.

As previously introduced in Sec.~\ref{subsec:onsourceregions}, an external EM or neutrino 
trigger at time $t_{0}$ can be used to define an
astrophysically motivated on-source window, such that the expected
GW counterpart associated with the external trigger is enclosed
within the on-source window. For the purposes of this study, we choose four distinct
on-source windows centered about $t_{0}$ -- see
Sec.~\ref{subsec:onsourceregions} for detailed information. Given a specified 
external source location, $(\alpha,\delta)$, the $N$ data streams observed from 
an $N$-detector network are time-shifted, such that any GW signals present 
will arrive simultaneously in each detector.  The time-shifted data streams are then 
projected onto the dominant polarization frame, in which GW signals
are maximized, and null frame, in which GW signals do not exist by
construction~\cite{klimenko:05,klimenko:06}.  

The data streams in the dominant polarization frame are processed to construct 
spectrograms, and the 1\% of time-frequency pixels with the largest
amplitude are marked as candidate signal events. For each cluster, a 
variety of information on the time and frequency characteristics is computed, in addition 
to measures of cluster significance, which are dependent on the total strain
energy $|h|^{2}$, of the cluster.
For the purposes of this study, a Bayesian likelihood 
statistic is used to rank the clusters. We direct the reader to~\cite{sutton:10,was:12} for a
detailed discussion of the cluster quantities used by \texttt{X-Pipeline}.

For statements on the detection of GWs to be made, we must be able to show with
high confidence that candidate events are statistically inconsistent with the
background data. To do this, we consider the
loudest event statistic, where the loudest event is the cluster in the
the on-source with the largest significance; we hereafter denote the
significance of the loudest event
$\mathcal{S}_{\mathrm{max}}^{\mathrm{on}}$~\cite{brady:04,biswas:09}. We
estimate the cumulative distribution of the loudest significances of background events,
$\mathcal{C}(\mathrm{S}_{\mathrm{max}})$, and set a threshold on the false alarm
probability (FAP) that the background could produce an event cluster in the 
on-source with significance $\mathcal{S}_{\mathrm{max}}^{\mathrm{on}}$. If
$\mathcal{C}(\mathrm{S}^{\mathrm{on}}_{\mathrm{max}})$ is greater than the
threshold imposed, we admit the loudest event as a potential GW detection
candidate. For the purposes of this study, we impose FAP = 0.1\%, which
corresponds to $\sim3.3\sigma$ confidence.

For Gaussian noise, the significance distribution of background events can be
estimated analytically, but as mentioned in Sec.~\ref{subsec:glitches}, glitches 
produce excess-power clusters in the data that may be mistaken for a GW event. 
However, the method used by 
\texttt{X-Pipeline} to construct the dominant polarization frame results in strong
correlations between the incoherent energy $I$ (from the individual data
streams) and the coherent energy $E$ (from the combined data streams) for
glitches~\cite{chatterji:06}. A comparison of $I$ and $E$ for candidate events 
can thus be used to \emph{veto} events that have the same statistical properties 
as the background noise. A threshold curve in $(I,E)$ space is defined, and veto 
tests may be one-sided (all events on one side of the curve are vetoed), or 
two-sided (events within some band centered on the $I = E$ diagonal are 
vetoed). The threshold curve is chosen to optimize the ratio of glitch rejection 
to signal acceptance. 

In practice, the statistics of the distribution of background events in the data are
determined by applying unphysically large time-shifts, hereafter referred to as
``lags'', to the detector streams. Additionally, we generate known 
signal events by injecting simulated GW signals into the data streams.  The
background and signal events are split into two sets, used for pipeline
\emph{tuning} and testing detection performance, respectively.  A large range of
trial threshold cuts are applied for the background rejection test, and the
statistics of the background events computed.  The minimum injection amplitude
for which $50\%$ of the injections (1) survive the threshold cuts and (2)
have a FAP $\leq 0.1\%$, $\hrssf$, for a given family of GW signal models is computed.
This is known as the upper limit on $\hrss$ at $50\%$
confidence -- see Sec.~\ref{subsec:upperlimits}. The optimal threshold 
cut is defined as that for which $h_\mathrm{rss}^{50\%}$ is minimized at the 
specified FAP. Unbiased statements on the 
background distribution and waveform detectability can then be made by processing 
the tuning set events with the thresholds obtained previously.

\subsection{Recoloring of GW detector data}
\label{subsec:recolor}
The many methods used to detect GW transients can often be proven to be
near optimal 
in the case of stationary, Gaussian noise. Data from the GW detectors, however, 
is not expected to be stationary or Gaussian, and as such, it is important to test the
efficacy of one's detection method in non stationary and non-Gaussan noise. To
this end, we utilize observational data taken by the 
Hanford and Livingston LIGO detectors during the S5 science run, in addition to data
taken by the Virgo detector during the VSR1 science run. The S5 data is now publically
available via the LIGO Open Science Center (LOSC)~\cite{vallisneri:15}.
Recoloring of these data to the predicted power spectral densities (PSDs) of 
the Advanced detectors during different stages during the next five years (see
Sec.~\ref{subsec:detectornoise}) permits a more realistic estimation of the
sensitivity of the advanced detectors to CCSNe.

We recolor the GW data using the \texttt{gstlal} software
packages~\cite{cannon:12,privitera:14}, following the procedure outlined below:
\begin{itemize}
\item Determine PSD of original data.
\item Whiten data using a zero-phase filter created from the original PSD.
\item Recolor whitened data to desired PSD.
\end{itemize}
This method provides non-Gaussian, nonstationary detector data including noise
transients, tuned to any sensitivity desired. For specific details on the detector 
networks, and noise PSDs considered, see Sec.~\ref{subsec:detectornoise}. For
the purposes of this study, we recolor 100 hours of data from the
H1 and L1 detectors during the S5 science run, and the V1 detector during the
VSR1 science run.


\subsection{Injection of known signal events}
\label{subsec:injections}
As mentioned previously in Sec.~\ref{subsec:x}, it is a well-established
practice to inject known signal events
into detector data for analysis (see, e.g.,~\cite{S6Burst}). This process 
permits the estimation of detection efficiency for GWs from signal models of 
varying time-frequency characteristics.

A GW source can be characterized by five angles--
$(\iota,\phi;\theta,\Phi,\psi)$, where $(\theta,\Phi,\psi)$ describe the sky
location and polarization of the source, while $(\iota,\phi)$ describe the
internal orientation of the source relative to the line of sight of the
observer. In this study, the source location in Earth-centered coordinates 
$(\theta,\Phi)$ are fixed by right ascension $\alpha$, and declination
$\delta$ of the source, in addition to the GPS time at geocenter of the
injected signal--see Sec.~\ref{sec:obs_scenarios} for more detailed information.
The polarization angle $\psi$ relating the source and detector reference frames
is distributed uniformly in $[0,2\pi]$ for all injections.  For CCSN systems,
it is not possible to know the inclination angle $\iota$ and
azimuthal angle $\phi$.  To represent this, we inject signals with many
different $(\iota,\phi)$, to average over all possible internal source
orientations. 

As mentioned previously in Sec.~\ref{sec:models}, we may construct the 
strain for different internal source orientations by projecting the mode
coefficients $\hlm$ onto the $-2$ spin-weighted spherical harmonics, $\sylm$.
Making use of geometric symmetries for different astrophysical systems permits 
the use of polarization factors to describe $h_{+,\times}(\iota,\phi)$ as
a function of $h_{+,\times; 0} = h_{+,\times}(\iota=0,\phi=0)$.
Defining polarization
factors $n_{+,\times}(\iota,\phi)$, we may write the strain at an arbitrary internal
orientation as
\begin{align}
h_{+}(\iota,\phi) &= n_{+}(\iota,\phi)h_{+;0}\,,\\
h_{\times}(\iota,\phi) &= n_{\times}(\iota,\phi)h_{\times;0}\,,
\end{align}
where the form of $n_{+,\times}(\iota,\phi)$ is dependent on the symmetries of
the system considered.

For linearly polarized signals (e.g., linear sine-Gaussian injections), we
apply
\begin{align}
n_{+}^{\mathrm{lin}} &= 1\,,\\
n_{\times}^{\mathrm{lin}} &= 0\,.
\end{align}

For elliptically polarized signals (e.g., bar-mode instability, disk
fragmentation instability, and elliptical sine-Gaussian injections), we
apply
\begin{align}
n_{+}^{\mathrm{el}} &= \frac{1}{2}\left(1+\cos\iota\right)^{2}\,,\\
n_{\times}^{\mathrm{el}} &= \cos\iota\,.
\end{align}

For the 2D CCSN emission models, the axisymmetric system results in a linearly
polarized GW signal.  The system has azimuthal symmetry, resulting in zero
amplitude for all GW modes except $H_{20}$. From Eq.~(\ref{eq:mode_expansion}),
we see that the strain $h_{+}$ varies with $\iota$ as 
\begin{align*}
h_{+}(\iota) &= h_{+}^{\mathrm{eq}}\,\sin^{2}\iota\,,
\end{align*}  
where $h_{+}^{\mathrm{eq}}$ is the strain as seen by an equatorial observer.  We
are thus able to apply \texttt{SN} polarization factors
\begin{align*}
n_{+}^{\mathrm{SN}} &= \sin^{2}\iota\,,\\
n_{\times}^{\mathrm{SN}} &= 0.
\end{align*}

For the 3D CCSN emission models, the GW polarizations are nontrivially
related to the internal source angles, and as such, the $h_{+}$ and $h_{\times}$
strains must be computed for specific internal configurations using
Eq.~\ref{eq:mode_expansion}. No additional polarization factors are applied for these
waveforms.

For all emission models for which $n_{+,\times}$ can be defined, we inject signals
uniform in $\cos\iota \in [-1,1]$.  For the $3D$ CCSN emission models, we
inject signals uniformly drawn from a bank of 100 realizations of
$(\cos\iota,\phi)$, where $\cos\iota \in [-1,-7/9,\ldots,1]$ and $\phi \in
[0,2\pi/9,\ldots,2\pi]$.

For each observational scenario, we inject $250$ injections across the
considered on-source window.

\subsection{Upper limits and detection efficiencies}
\label{subsec:upperlimits}
To make detection statements and set upper limits on the GWs emitted from CCSNe,
we must compare the cumulative distribution of background event significance, 
$\mathcal{C}(\mathcal{S}_{\mathrm{max}})$, estimated from off-source 
data, to the maximum event significance in the on-source data
$\mathcal{S}_{\mathrm{max}}^{\mathrm{on}}$. If no on-source events are
significant, we may instead proceed to set frequentist upper limits on the GWs
from the CCSN of interest, given the emission models considered.

As alluded to previously in Sec.~\ref{subsec:x}, we may define the $50\%$
confidence level upper limit on the signal amplitude for a specific GW emission
model as the minimum amplitude for which the probability of observing the
signal, if present in the data, with a cluster significance louder than
$\mathcal{S}_{\mathrm{on}}^{\mathrm{max}}$ is $50\%$. In this study, we aim to
determine the $50\%$ upper limit, as defined here, as a function of
\begin{itemize}
\item Source distance $d^{50\%}$, in the context of astrophysically motivated
signal models.
\item Root-sum-square amplitude $h_{\mathrm{rss}}^{50\%}$, in the context of
linear and elliptical sine-Gaussian waveforms. It is more relevant,
astrophysically to consider the corresponding $50\%$ upper limit on the energy
emitted in GWs, $E^{50\%}_{\mathrm{GW}}$, which we compute from
$h_{\mathrm{rss}}^{50\%}$ using Eq.~(\ref{eq:egw_sg}).
\end{itemize} 

After the on-source data has been analyzed and
$\mathcal{S}_{\mathrm{on}}^{\mathrm{max}}$ computed, we inject a large
number of known signal events for families of
waveforms for which $h_{\mathrm{rss}}^{50\%}$ and $d^{50\%}$ (where applicable)
are desired. For a single waveform
family, we outline the upper limit procedure:
\begin{itemize}
\item Inject many waveforms at different times during the on-source window and with a
broad range of polarization factors.
\item Compute the largest significance $\mathcal{S}$ of any clusters associated 
with the injected waveforms (observed within ~0.1 seconds of the injection time) 
that have survived after application of veto cuts.
\item For all injections, compute the percentage of injections for which
$\mathcal{S} > \mathcal{S}_{\mathrm{on}}^{\mathrm{max}}$. This is called the
detection efficiency, $\mathcal{E}$.
\item Repeat procedure, modifying the injection amplitude of each waveform by
a scaling factor.  
\end{itemize}
The final goal is to produce a plot of the detection efficiency as a
function of $h_{\mathrm{rss}}$ or distance $d$ for each waveform family, such
that one may place upper limits on the GW emission models considered.  From the
efficiency curve, one may determine $h_{\mathrm{rss}}^{50\%}$ as 
\begin{align}
\mathcal{E}(h_{\mathrm{rss}} = h_{\mathrm{rss}}^{50\%}) = 0.5\,. 
\end{align}
Given an astrophysical signal injected at $h^{\mathrm{inj}}_{\mathrm{rss}}$ 
corresponding to fiducial distance $d^{\mathrm{inj}}$, we may define
$d^{50\%}$ as
\begin{align}
d^{50\%} &=
\left(\frac{h_{\mathrm{rss}}^{50\%}}{h^{\mathrm{inj}}_{\mathrm{rss}}}\right)d^{\mathrm{inj}}\,.
\end{align}
We note that \texttt{X-Pipeline} rescales the detection efficiency to account
only for injections placed at times at which detector data is available. Without
this correction, the efficiencies computed asymptote to the duty cycle fraction
for the on-source window considered. For the data considered in this study, the
total duty cycle is typical of the S5 and VSR1 science runs, which is described
in detail in Sec.~\ref{subsec:dutycycle}.

\subsection{Systematical uncertainties}
The uncertainties in the efficiencies, upper limits and exclusion capabilities
of our analysis method are related to non-Gaussian transients in the data, in
addition to calibration uncertainties. There are a number of systematic
uncertainties present in this study that will non-negligibly affect the results.
We consider only a short period of recolored data from LIGO's S5 and Virgo's
VSR1 data-taking runs, over which the frequency and character of 
non-Gaussian transients changed non-negligibly. The noise transients in advanced
LIGO data are also significantly different to those in initial LIGO data, and
the non-Gaussianities are not yet understood well enough to make quantitative
statements on the statistical behavior of the data. For these reasons, we only
quote results to two significant figures in this study.
The statistical uncertainty in detector calibration can be
characterized by the $1\sigma$ statistical uncertainty in the amplitude and
phase of the signal.  Uncertainties in phase calibration can be
estimated by simulating its effect on the ability to recover test injections.
We direct the reader to Kalmus~\cite{kalmus:09}, in which it is shown that 
phase uncertainties contribute negligibly to the total systematic 
error, and thus we only consider amplitude uncertainties in this study. The
target design amplitude uncertainties in the frequency range 
40-2048\,Hz for Advanced LIGO and Advanced Virgo are 
$5\%$ at $2\sigma$ confidence~\cite{aLIGO:15}. As such, the upper limits for
$h_{\mathrm{rss}}^{50\%}$ and $d^{50\%}$ obtained from a search for GWs from
CCSNe in the Advanced detector era will have intrinsic $\sim5\%$
uncertainties. For comparison, typical amplitude uncertainties due to
calibration in S5 were below $15\%$~\cite{abadie:10_s5calib}.

\section{Results}
\label{sec:results}
In this section, we present the results for the detectability of the considered
GW emission models described in Sec.~\ref{sec:models}. 

We consider realistic
waveform models from numerical simulations of core collapse. For the
`garden-variety' CCSN models considered
(\texttt{m\"uller1}, \texttt{m\"uller2}, \texttt{m\"uller3},
\texttt{ott}, and \texttt{yak}), convection and SASI are the dominant GW emission
processes. For rotating core collapse, we choose models where bounce and ringdown of
the proto-neutron star (\texttt{dim1}, \texttt{dim2}, and \texttt{dim3}), and
nonaxisymmetric rotational instabilities (\texttt{sch1} and \texttt{sch2}) are
the dominant GW emission processes. As these waveforms will only be detectable from
CCSNe at close distances ($d \lesssim 100\,\mathrm{kpc}$), we consider CCSNe in 
the direction of the Galactic center and LMC, for which the coincident neutrino
signal will be detected. We use a conservative on-source window of [-10,+50]s 
about the time of the initial SNEWS trigger.

For more distant CCSNe, we consider more speculative, extreme phenomenological
GW emission models for long-lived bar-mode instabilities (\texttt{longbar1},
\texttt{longbar2}, \texttt{longbar3}, \texttt{longbar4}, \texttt{longbar5}, and
\texttt{longbar6}) and disk fragmentation instabilities (\texttt{piro1}, 
\texttt{piro2}, \texttt{piro3}, and \texttt{piro4}). More distant CCSNe will not
be detectable via neutrinos, but the EM counterpart will be observed. We
consider CCSNe in M31 and M82, and use on-source windows assuming a compact,
stripped progenitor star of 61 minutes and 24 hour 1 minute, respectively. For
an extended, red supergiant progenitor, we use on-source windows of 51 hours and
74 hours for M31 and M82, respectively.

For all host galaxies, we consider \emph{ad hoc} sine-Gaussian bursts to assess
the sensitivity of our analysis to localized bursts of energy in time-frequency
space.

We remind the reader of the large systematic uncertainties associated with these
results and, as such, quote all results to two significant figures.

\subsection{Numerical waveforms}
\label{subsec:numwaveform_results}
We present the distances $d^{50\%}$ at which 50\% detection efficiency is attained
(the measure we use for ``detectability'') for the considered 
numerical waveforms in Table~\ref{tab:numerical_d50},
for CCSNe in the direction of the Galactic center and LMC, in the context
of a 60-second on-source window.

\begin{table*}
\centering
\begin{tabular}{c|rrr|rrr}
\multicolumn{1}{c}{} &
\multicolumn{3}{c}{$d^{50\%}$ $[\mathrm{kpc}]$ for Galactic center} &
\multicolumn{3}{c}{$d^{50\%}$ $[\mathrm{kpc}]$ for LMC} \\
Waveform & \texttt{HL 2015} & \texttt{HLV 2017} &
\texttt{HLV 2019}&\texttt{HL 2015} & \texttt{HLV 2017} &
\texttt{HLV 2019}\\
\hline
\hline
\texttt{m\"{u}ller1} & 2.3 & 3.3 & 4.7    & 2.5 & 3.8 & 5.3  \\
\texttt{m\"{u}ller2} & 1.0 & 1.5 & 2.2    & 1.2 & 1.8 & 2.5  \\
\texttt{m\"{u}ller3} & 1.2 & 1.5 & 2.4    & 1.4 & 1.6 & 2.7  \\
\texttt{ott}         & 2.4 & 3.4 & 5.5    & 3.2 & 4.9 & 7.2  \\
\texttt{yak}         & 1.5 & 1.8 & 5.1    & 1.6 & 2.1 & 6.2  \\
\texttt{dim1} 	     & 7.0 & 9.1 & 17\phantom{0}   & 7.4 & 10\phantom{0} & 18\phantom{0} \\
\texttt{dim2} 	     & 11\phantom{0} & 17\phantom{0} & 29\phantom{0} & 13\phantom{0} & 20\phantom{0} & 32\phantom{0} \\
\texttt{dim3} 	     & 13\phantom{0} & 21\phantom{0} & 38\phantom{0} & 18\phantom{0} & 32\phantom{0} & 50\phantom{0} \\
\texttt{sch1}        & 31\phantom{0} & 43\phantom{0} & 78\phantom{0} & 36\phantom{0} & 48\phantom{0} & 90\phantom{0}  \\
\texttt{sch2}        & 35\phantom{0} & 50\phantom{0} & 98\phantom{0} & 45\phantom{0} & 56\phantom{0} &120\phantom{0}  \\
\hline
\hline
\end{tabular}
\caption{The distance in kpc at which 50\% detection efficiency is attained,
$d^{50\%}$ for the numerical core-collapse emission models considered 
using the \texttt{HL 2015}, \texttt{HLV 2017}, and 
\texttt{HLV 2019} detector networks, for CCSNe in the direction of the Galactic
center and the LMC.}
\label{tab:numerical_d50}
\end{table*}

For CCSNe in the direction of the Galactic center, we see that emission from 
neutrino-driven convection and SASI is detectable out to 
$\sim$\,(1.0-2.4)\,kpc with the
\texttt{HL 2015} detector network. This increases to
$\sim$\,(1.5-3.4)\,kpc and $\sim$\,(2.2-5.5)\,kpc with the
\texttt{HLV 2017} and \texttt{HLV 2019} detector networks, respectively.

Similarly, we see that emission from bounce and ringdown of the 
central proto-neutron star core is detectable out to 
$\sim$\,(7.0-13.4)\,kpc for CCSNe in the direction of the Galactic center
with the \texttt{HL 2015} detector network. This increases to
$\sim$\,(9.1-21)\,kpc and $\sim$\,(17-38)\,kpc with the
\texttt{HLV 2017} and \texttt{HLV 2019} detector networks, respectively.

Emission from nonaxisymmetric rotational instabilities from CCSNe in the
direction of the galactic center is detectable out to
$\sim$\,(31-35)\,kpc with the \texttt{HL 2015} detector network. This
increases to $\sim$\,(43-50)\,kpc and $\sim$\,(78-98)\,kpc
with the \texttt{HLV 2017} and \texttt{HLV 2019} detector networks,
respectively. 

Assuming the fiducial distance of a galactic CCSN to be $\sim9\,\mathrm{kpc}$,
this suggests that we will be able to detect emission from the more extremely
rapidly rotating CCSN waveforms considered with the \texttt{HL 2015} detector
network, while all considered rapidly rotating waveforms will be detectable for
CCSNe in the direction of the Galactic center 
with the \texttt{HLV 2017} and \texttt{HLV 2019} detector networks. We will be
limited to detection of nonrotating CCSNe within $5.5\,\mathrm{kpc}$ with the
most sensitive \texttt{HLV 2019} detector network.

Considering CCSNe in the direction of the LMC, we see that emission from 
neutrino-driven convection and SASI is detectable out to 
$\sim$\,(1.2-3.2)\,kpc with the
\texttt{HL 2015} detector network. This increases to
$\sim$\,(1.6-4.9)\,kpc and $\sim$\,(2.5-7.2)\,kpc with the
\texttt{HLV 2017} and \texttt{HLV 2019} detector networks, respectively. Given
that the LMC is $\sim50\,\mathrm{kpc}$ away, this
shows that emission from neutrino-driven convection and SASI will 
not be detectable from CCSNe in the LMC. 

Emission from bounce and ringdown of the 
central proto-neutron star core is detectable out to 
$\sim$\,(7.4-18)\,kpc and $\sim$\,(11-32)\,kpc for CCSNe in the 
direction of the LMC with the \texttt{HL 2015} and \texttt{HLV 2017} detector
networks, respectively. This increases to
$\sim$\,(18-50)\,kpc with the \texttt{HLV 2019} detector network.
This suggests that emission from the bounce and subsequent
ringdown of the proto-neutron star may not be detectable from CCSNe in the LMC
for even the most rapidly rotating waveform considered with the \texttt{HLV
2019} detector network. 

We see that emission from nonaxisymmetric 
rotational instabilities from CCSNe in the direction of the LMC is detectable out to
$\sim$\,(36-45)\,kpc with the \texttt{HL 2015} detector network. This
increases to $\sim$\,(48-56)\,kpc and $\sim$\,(90-120)\,kpc
with the \texttt{HLV 2017} and \texttt{HLV 2019} detector networks,
respectively. This suggests we will be able to detect emission from
nonaxisymmetric rotational instabilities for CCSNe in the LMC
with the \texttt{HLV 2017} detector network. 

\begin{figure*}[!ht]
\includegraphics[width=\columnwidth]{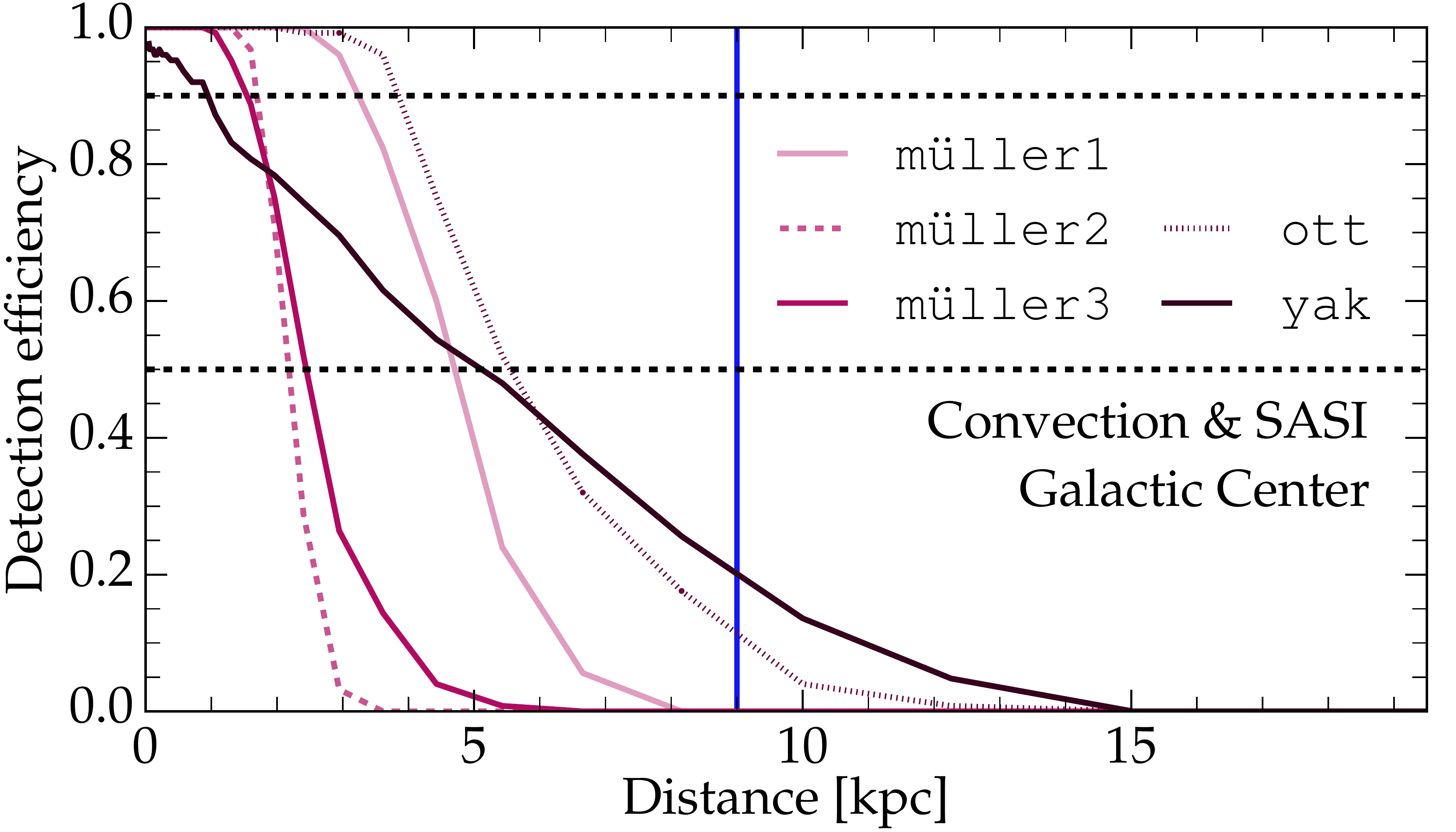}
\includegraphics[width=\columnwidth]{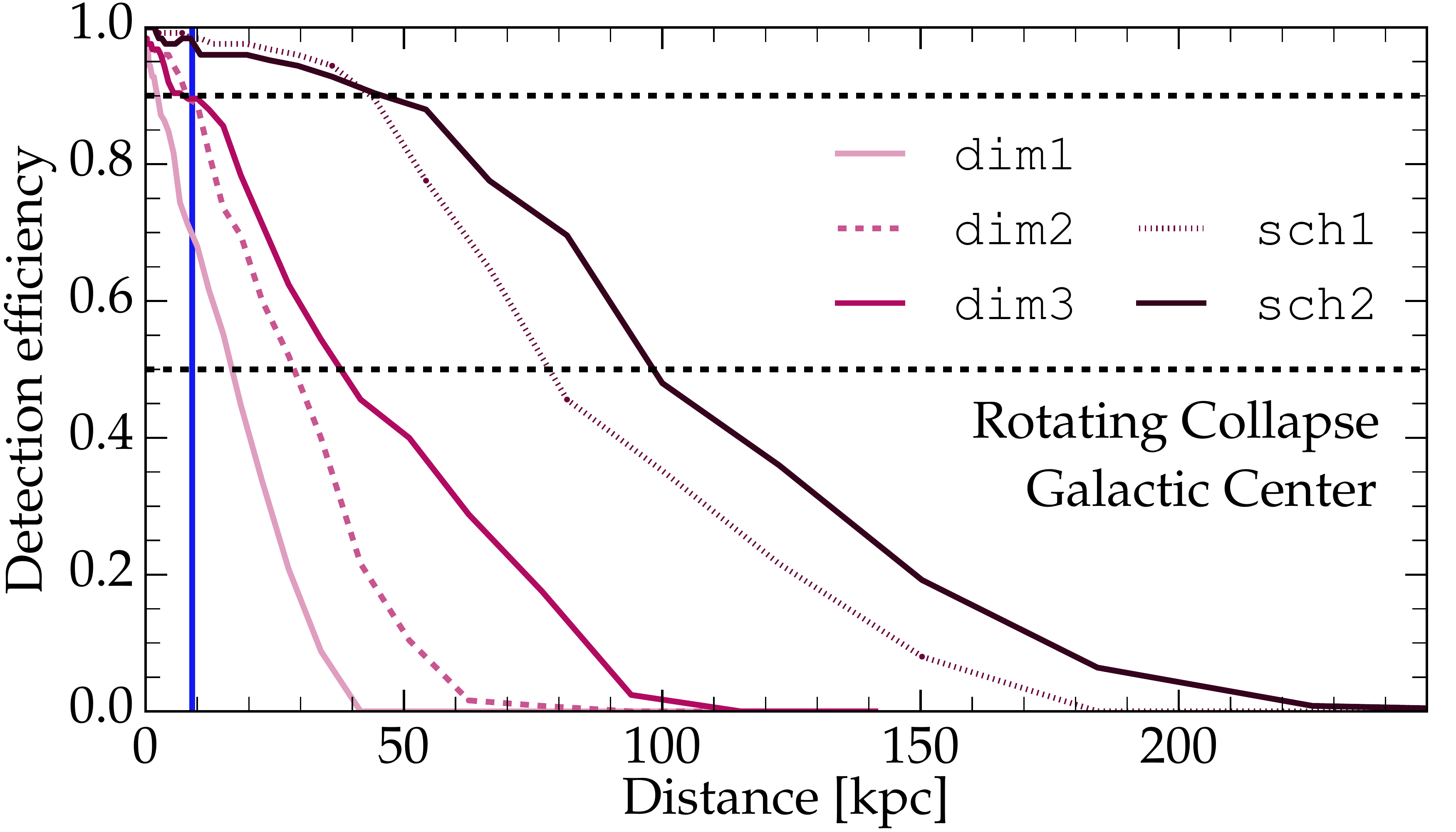}\\
\includegraphics[width=\columnwidth]{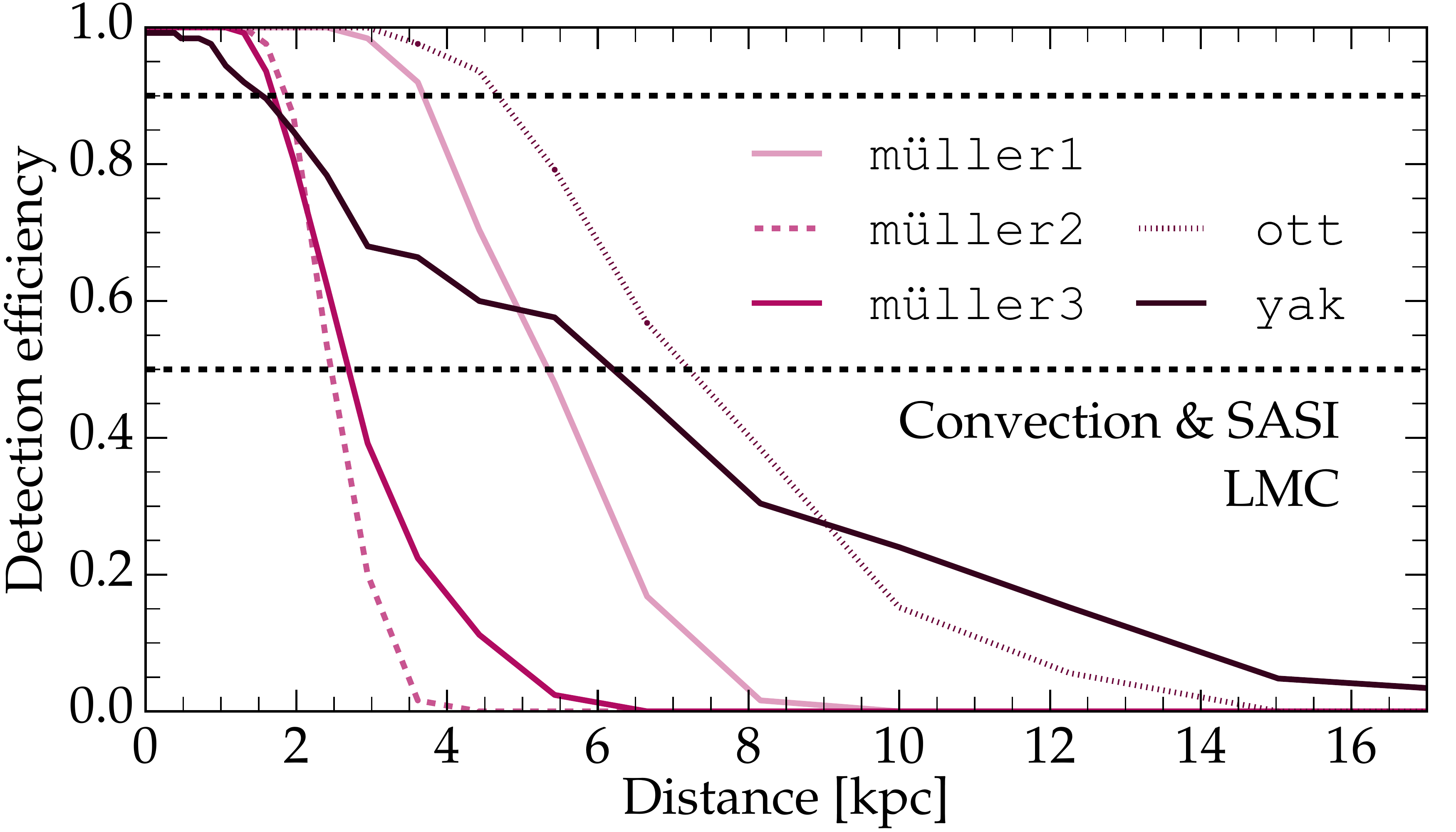}
\includegraphics[width=\columnwidth]{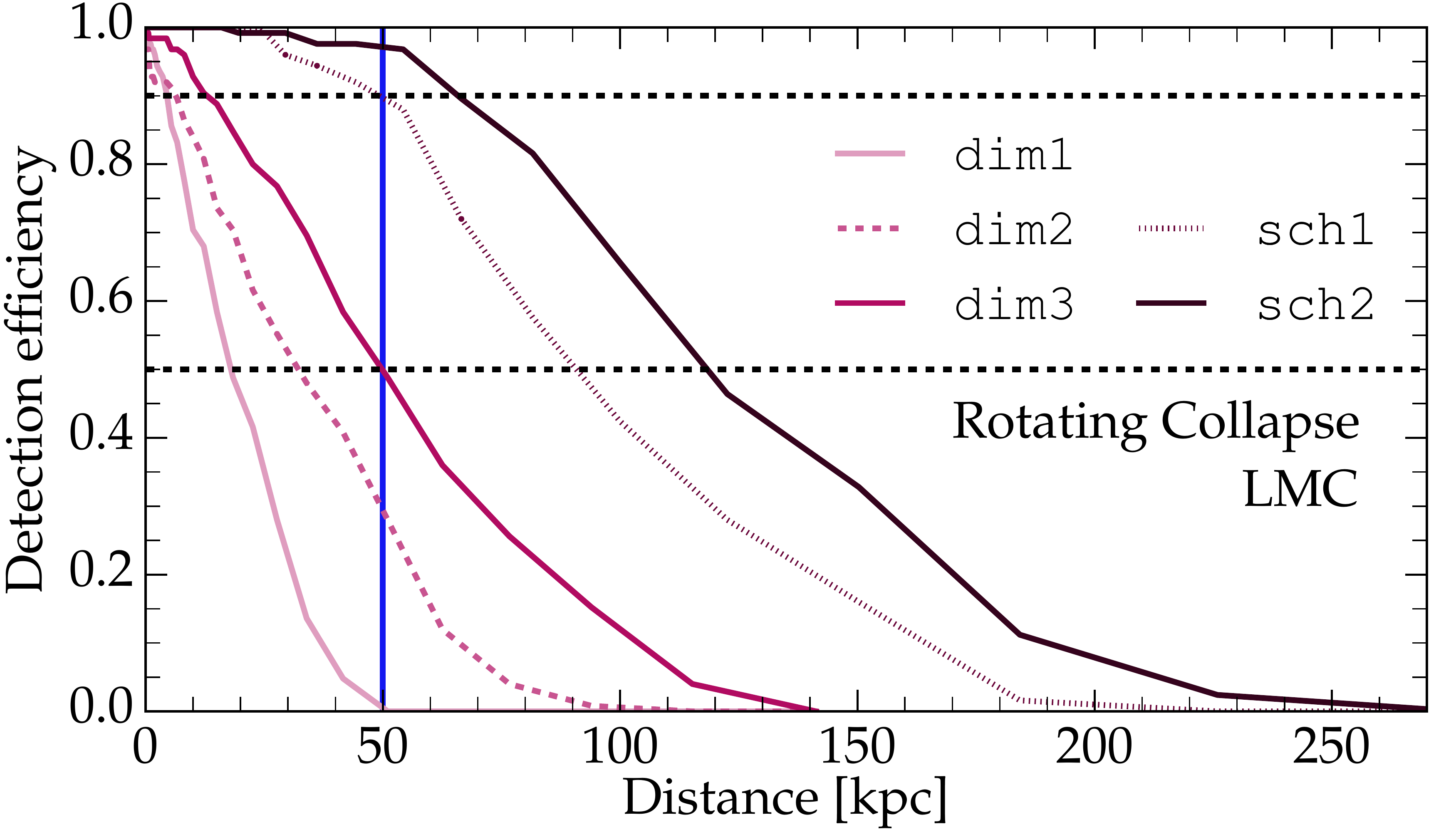}
\caption{The detection efficiency as a function of distance for the numerical
waveforms in this study, in the context of a 1 minute 
on-source window and the \texttt{HLV 2019} detector network. The top row is for 
galactic sources, and the bottom row is for sources in the Large Magellanic 
Cloud. In each plot, $50\%$ and $90\%$ detection efficiency is marked with a
dashed black line, and the distance to the host galaxy is marked with a vertical
blue line.}
\label{fig:MW-LMC_HLV2019}
\end{figure*}

Figure~\ref{fig:MW-LMC_HLV2019} presents the detection efficiency as a
function of distance, for the numerical waveforms considered, for CCSNe 
directed toward the Galactic center and the LMC.

\subsection{Extreme phenomenological models}
\label{subsec:phenomwaveform_results}
We present the distances at which 50\% detection efficiency is attained
$d^{50\%}$ (the measure we use for ``detectability'') for the considered 
phenomenological waveforms in Table~\ref{tab:phenom_d50},
for CCSNe in the direction of M31, in the context of 61-minute and 51-hour
on-source windows, and M82, in the context of 24-hour 1-minute and 74-hour
on-source windows.

\begin{table*}[!ht]
\centering
\begin{tabular*}{0.75\textwidth}{@{\extracolsep{\fill} }c|rrr|rrr}
\multicolumn{1}{c}{} &
\multicolumn{3}{c}{$d^{50\%}$ $[\mathrm{Mpc}]$ for M31} &
\multicolumn{3}{c}{$d^{50\%}$ $[\mathrm{Mpc}]$ for M82} \\
Waveform & \texttt{HL 2015} & \texttt{HLV 2017} &
\texttt{HLV 2019}& \texttt{HL 2015} & \texttt{HLV 2017} &
\texttt{HLV 2019}\\
\hline
\hline
\texttt{longbar1} & 0.5 [0.2] & 0.8 [0.3] & 1.6 [0.8] & 0.3 [0.4]  & 0.3 [0.4] & 1.0 [0.7] \\
\texttt{longbar2} & 1.5 [0.7] & 2.5 [0.9] & 4.8 [2.8] & 0.9 [1.1]  & 1.0 [1.2] & 3.0 [2.1] \\
\texttt{longbar3} & 1.0 [0.6] & 1.6 [0.6] & 3.6 [2.2] & 0.8 [0.8]  & 0.7 [0.8] & 2.4 [1.8] \\
\texttt{longbar4} & 2.0 [1.1] & 2.8 [1.2] & 6.0 [3.8] & 1.1 [1.5]  & 1.4 [1.5] & 3.9 [2.8] \\
\texttt{longbar5} & 5.2 [2.7] & 8.6 [3.4] & 18\phantom{0} [9.9]& 3.0 [4.3]  &
3.4 [5.2] & 9.7 [8.3] \\
\texttt{longbar6} & 2.1 [1.1] & 3.4 [1.1] & 6.7 [4.7] & 1.4 [1.9]  & 1.4 [1.7] & 4.4 [3.7] \\
\texttt{piro1} 	  & 0.9 [0.6] & 1.3 [0.6] & 2.0 [1.4] & 0.5 [0.7]  & 0.7 [0.8]& 1.3 [1.3] \\
\texttt{piro2} 	  & 3.9 [2.2] & 6.3 [2.6] & 9.4 [5.8] & 2.2 [3.2]  & 3.0 [3.8]& 5.7 [5.8]  \\
\texttt{piro3} 	  & 1.9 [1.3] & 3.4 [1.8] & 4.9 [3.7] & 1.1 [1.3]  & 1.5 [1.9]& 2.8 [3.1] \\
\texttt{piro4}    & 12\phantom{0} [6.5] & 19\phantom{0} [6.1] & 28\phantom{0}
[18]\phantom{0} & 6.4 [7.5] & 8.6 [9.5] & 16\phantom{0} [15]\phantom{0}\\
\hline
\hline
\end{tabular*}
\caption{The distance in Mpc at which 50\% detection efficiency is attained,
$d^{50\%}$ for the extreme phenomenological emission models considered 
using the \texttt{HL 2015}, \texttt{HLV 2017}, and 
\texttt{HLV 2019} detector networks, for CCSNe in the direction of M31 and
M82, in the context of 61-minute (51-hour) and 24-hour 1-minute (74 hour)
on-source windows, respectively.}
\label{tab:phenom_d50}
\end{table*}

For CCSNe in the direction of M31, we see that emission from long-lived 
bar-mode instabilities will be detectable out to 
$\sim$\,(0.5-5.2)\,Mpc [$\sim$\,(0.2-2.7)\,Mpc] when using a 61-minute 
[51-hour] on-source window, with the \texttt{HL 2015} detector network.
The distances at which 50\% detection efficiency is reached, $d^{50\%}$, 
increase to $\sim$\,(0.8-8.6)\,Mpc [$\sim$\,(0.3-3.4)\,Mpc] and 
$\sim$\,(1.6-18)\,Mpc [$\sim$\,(0.8-9.9)\,Mpc] when using a 61-minute 
[51-hour] on-source window, with the \texttt{HLV 2017} and \texttt{HLV
2019} detector networks, respectively.

Emission from disk fragmentation instabilities will be detectable 
out to $\sim$\,(0.9-12)\,Mpc [$\sim$\,(0.6-6.5)\,Mpc] and 
$\sim$\,(1.3-19)\,Mpc [$\sim$\,(0.6-6.1)\,Mpc] when using 61-minute 
[51-hour] on-source windows with the \texttt{HL 2015} and 
\texttt{HLV 2017} detector networks, respectively, for CCSNe in the 
direction of M31. These distances increase to
$\sim$\,(2-28)\,Mpc [$\sim$\,(1.4-18)\,Mpc] when using a 61-minute
[51-hour] on-source window, with the \texttt{HLV 2019} detector network.

Assuming a fiducial distance of $0.77\,\mathrm{Mpc}$ for a CCSN in M31, this
suggests that we will be able to detect emission from all considered long-lived 
bar-mode instability waveforms with the \texttt{HLV 2019} detector network,
while the detectable fraction of considered waveforms with the \texttt{HL 2015}
and \texttt{HLV 2017} detector networks is strongly dependent on the on-source
window length. Taking the 51-hour on-source window as the most pessimistic
scenario, $\sim50\%$ and $\sim67\%$ of the considered bar-mode instability
waveforms are detectable with the \texttt{HL 2015} and \texttt{HLV 2017}
detector networks, respectively.

Similarly, emission from the considered disk fragmentation instabilities
waveforms will be detectable for a CCSN in M31 with the \texttt{HLV 2019}
detector network for all considered on-source windows. For the 51-hour on-source
window, we see that $\sim75\%$ of the considered disk-fragmentation instability
waveforms are detectable with both the \texttt{HL 2015} and \texttt{HLV 2017}
detector networks.

We note that, for some models, the $d^{50\%}$ values computed for the M31
source, when using a 51-hour on-source window, are smaller for the \texttt{HLV
2017} detector network than the \texttt{HL 2015} network. While this might at
first seem counter-intuitive, this is due to the requirement for
coincident data between detectors to run a coherent analysis.
The lower sensitivity of the \texttt{HV} and \texttt{LV} detectors for
the data analyzed, compared with the sensitivity of the \texttt{HL} detectors,
reduces the effective total sensitivity of the network. We include the third
detector, however, as it increases the overall duty cycle of the network.

For CCSNe in the direction of M82, we see that emission from 
long-lived bar-mode instabilities will be
detectable out to $\sim$\,(0.3-3)\,Mpc [$\sim$\,(0.4-4.3)\,Mpc]
and $\sim$\,(0.3-3.4)\,Mpc [$\sim$\,(0.4-5.2)\,Mpc]
using a 24-hour 1-minute [74-hour] on-source window, with the \texttt{HL
2015} and \texttt{HLV 2017} detector networks. This increases to 
$\sim$\,(1-9.7)\,Mpc [$\sim$\,(0.7-8.3)\,Mpc] for a 24-hour 1-minute 
[74-hour] on-source window, with the \texttt{HLV 2019} detector network.

For emission from disk fragmentation instabilities for CCSNe in the direction of
M82, the distance reach is $\sim$\,(0.5-6.4)\,Mpc
[$\sim$\,(0.7-7.5)\,Mpc] when using a 24-hour 1-minute [74-hour]
on-source window with the \texttt{HL 2015} detector network. This increases to 
$\sim$\,(0.7-8.6)\,Mpc [$\sim$\,(0.8-9.5)\,Mpc] and
$\sim$\,(1.3-16)\,Mpc [$\sim$\,(1.3-15)\,Mpc] for the \texttt{HLV
2017} and \texttt{HLV 2019} detector networks, respectively.

Given a fiducial distance of $\sim3.52\,\mathrm{Mpc}$ for CCSNe in M82, we note
that only the most extreme waveform considered for both long-lived bar-mode
instabilities and disk fragmentation instabilities are detectable with the
\texttt{HL 2015} detector network. Of the considered long-lived bar-mode
instability waveforms, only the most extreme emission model is detectable with
the \texttt{HLV 2017} detector network, while $50\%$ of the waveforms will be
detectable with the \texttt{HLV 2019} detector network. For emission from
disk fragmentation instabilities, we see that only $50\%$ of the waveforms
considered will be detectable out to M82 with the \texttt{HLV 2017} and
\texttt{HLV 2019} detector networks.

We note that the distance reach for these models increases with the larger
on-source window for the M82 source. This is due to the properties of the data
over the two considered on-source windows. As previously mentioned, real data
from GW detectors is not stationary, and as such, the PSD of the data is a
function of time. Time periods over which the detector data is glitchy will have
locally have significantly decreased sensitivity when compared to a much larger
time period over which the detector is more well behaved. This means that if the
on-source window derived happens to lie in a glitchy period of detector data,
the sensitivity of the detector network will, unfortunately, be decreased. In
repeating the search for a larger on-source window, over which the average
sensitivity is much greater, the distance reach for the emission models
considered may appear to increase. The detectability of the
waveforms considered in this study is established by injecting a number of
waveforms over the full on-source window considered. The distance reach for the
longer on-source window in this case appears to increase because we inject
waveforms uniformly across the on-source window, meaning that many ``test''
signals are placed at times in the data stretch where the sensitivity is
greater, in addition to the shorter, more glitchy, time period where the
sensitivity is not as great. This is a great example of how realistic noise 
can significantly affect the detectability of GWs from CCSNe at different times,
and is motivation for improving active noise suppression techniques for the
detectors.

\begin{figure*}[!ht]
\includegraphics[width=\columnwidth]{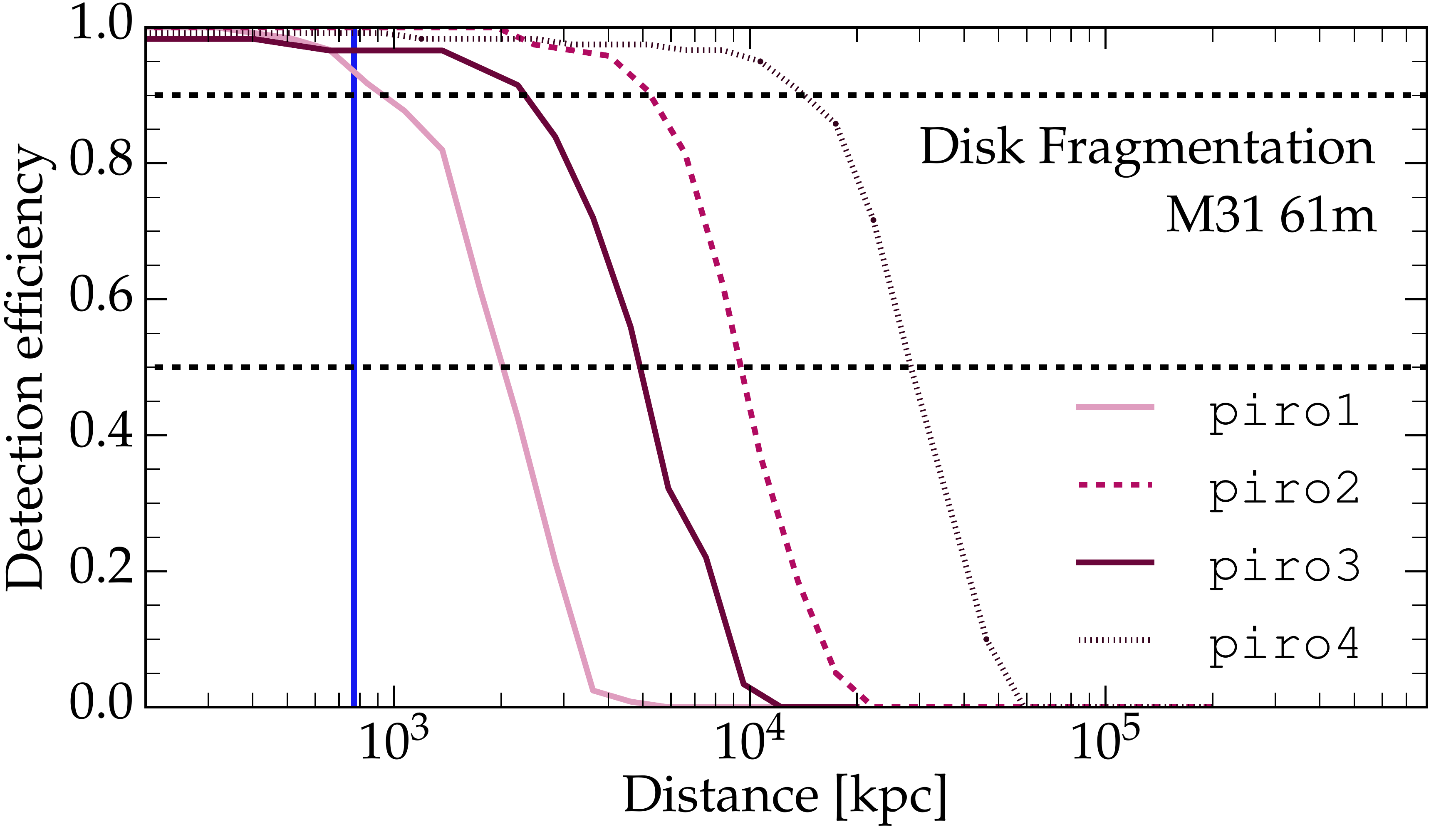}
\includegraphics[width=\columnwidth]{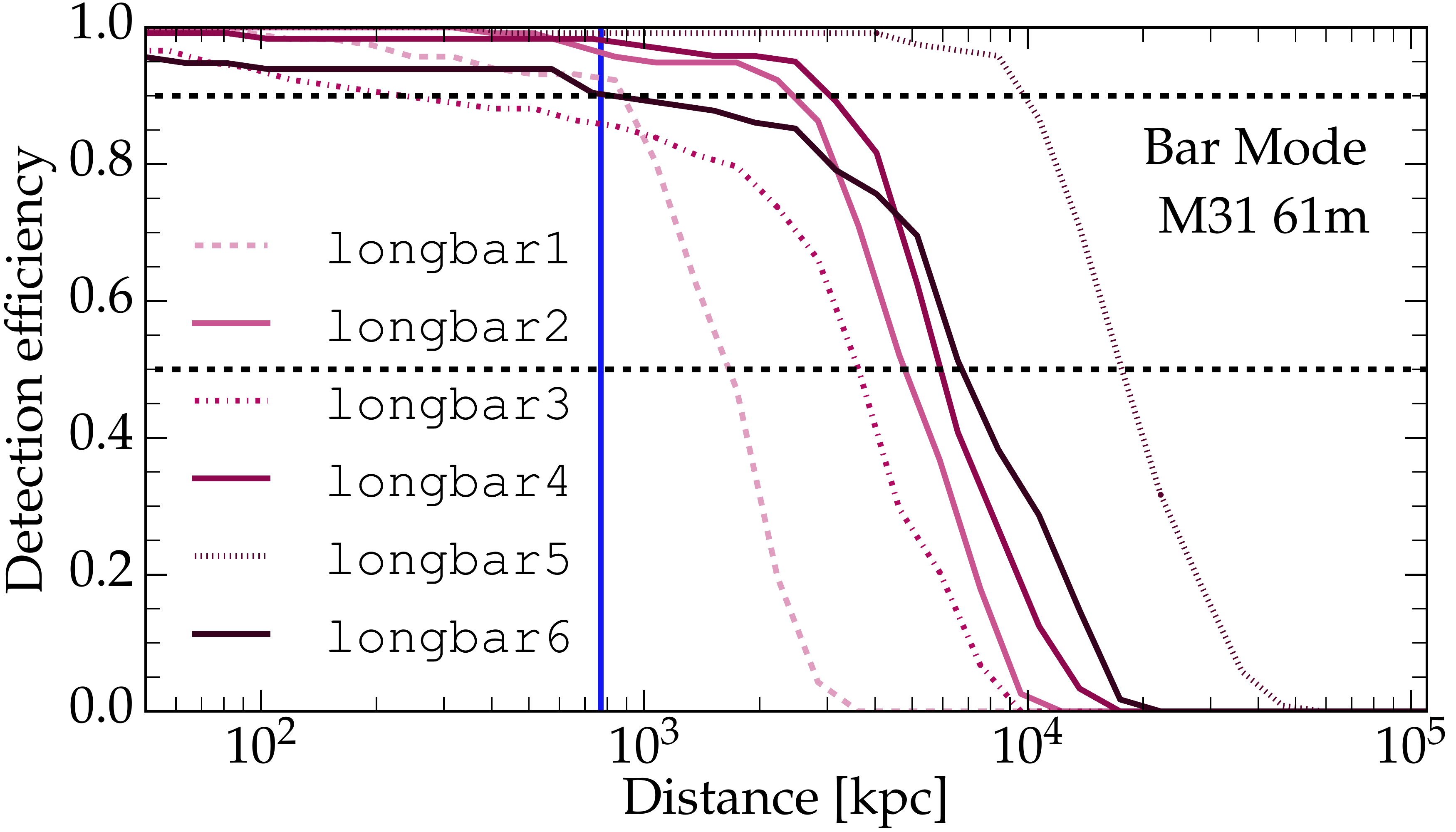}\\
\includegraphics[width=\columnwidth]{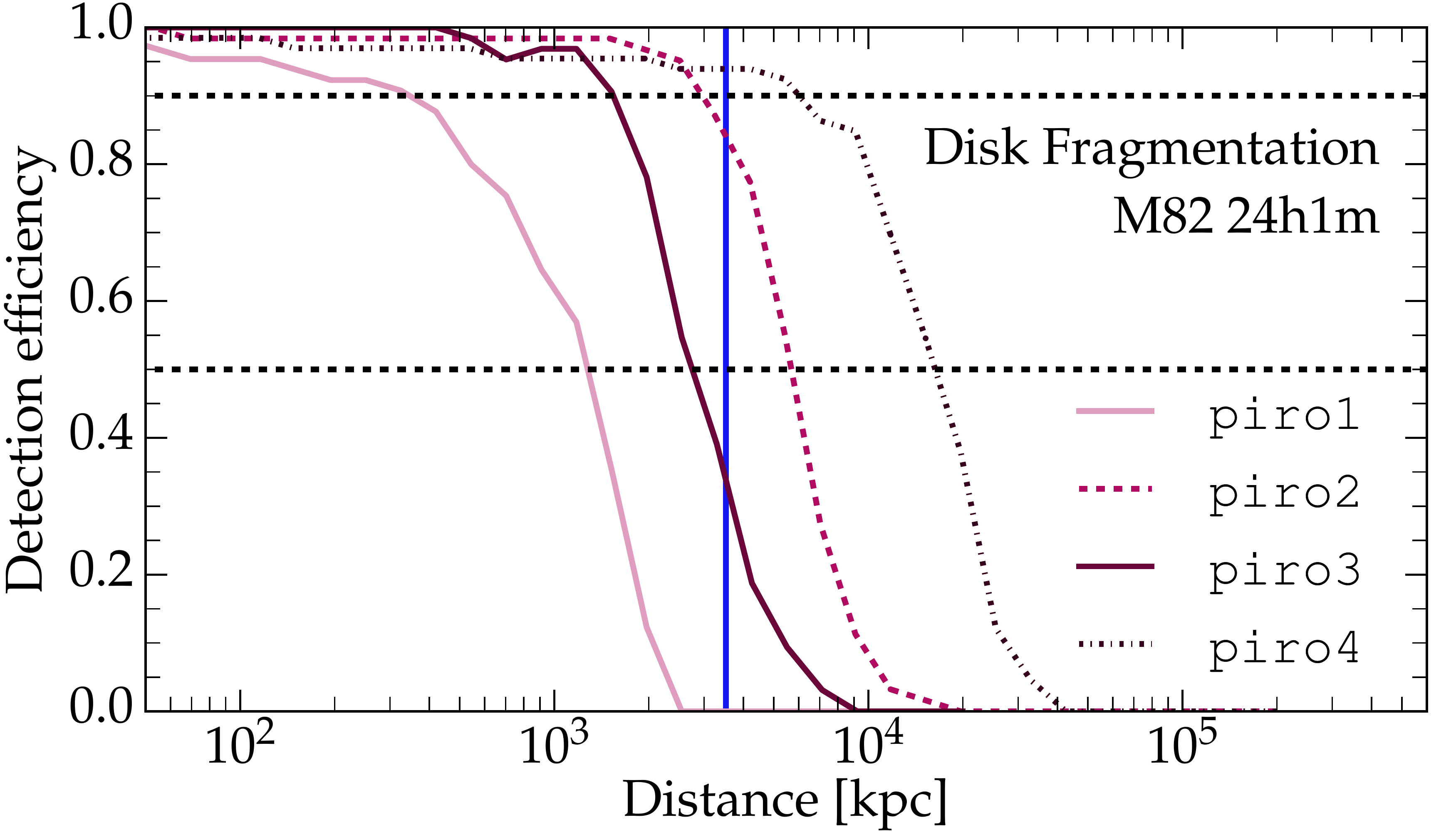}
\includegraphics[width=\columnwidth]{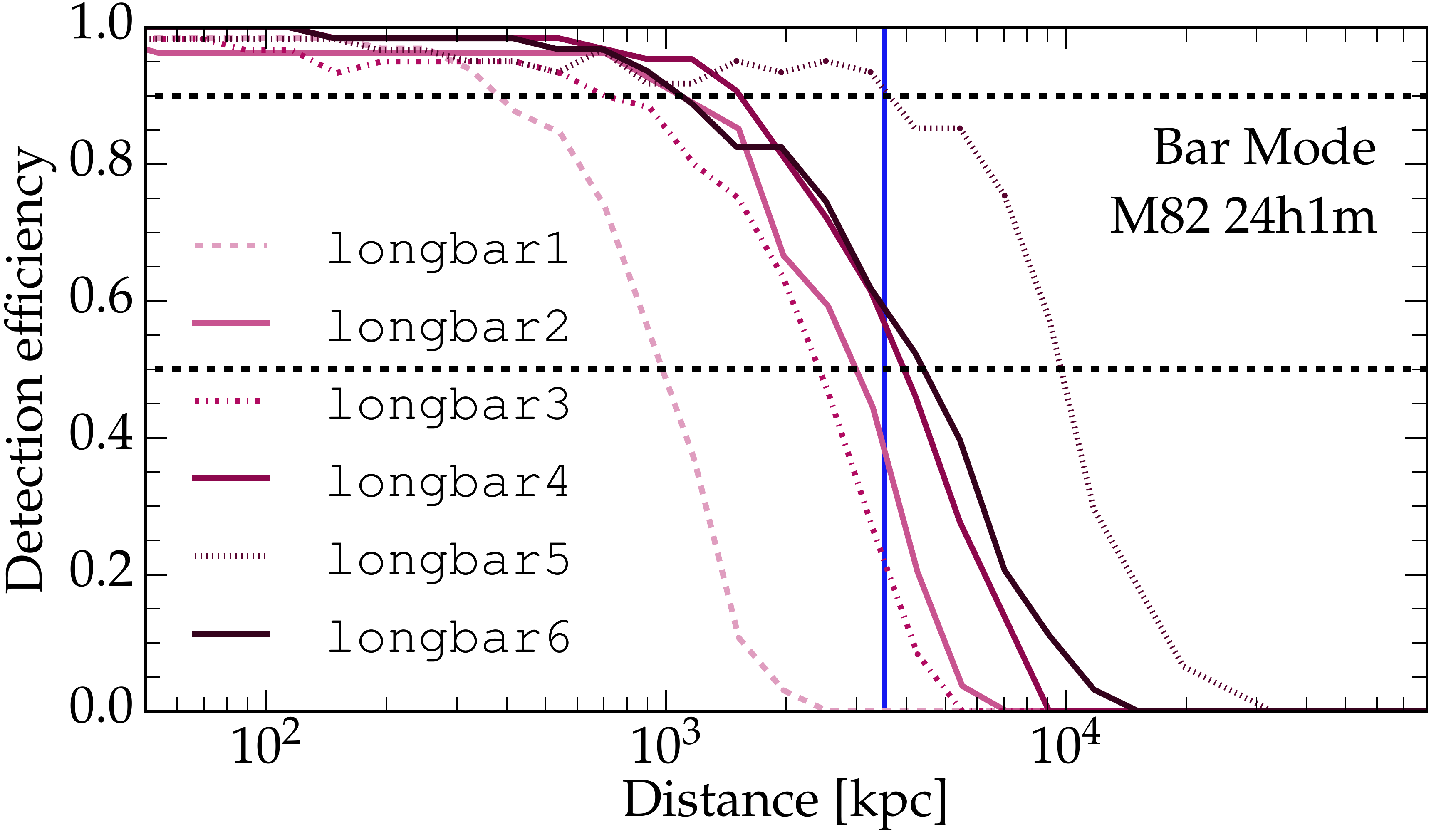}
\caption{The detection efficiency as a function of distance for the
phenomenological waveforms considered in this study, in the context of the 
on-source window astrophysically motivated by a stripped envelope type Ibc SN
progenitor and the \texttt{HLV 2019} detector configuration. The top row is for
sources in M31 with an on-source window of 61 minutes, and the bottom row is for
sources in M82 with a 24-hour 1-minute on-source window. In each plot, 
$50\%$ and $90\%$ detection efficiency is marked with a
dashed black line, and the distance to the host galaxy is marked with a vertical
blue line.}
\label{fig:M31-M82_HLV2019A}
\end{figure*}

\begin{figure*}[!ht]
\includegraphics[width=\columnwidth]{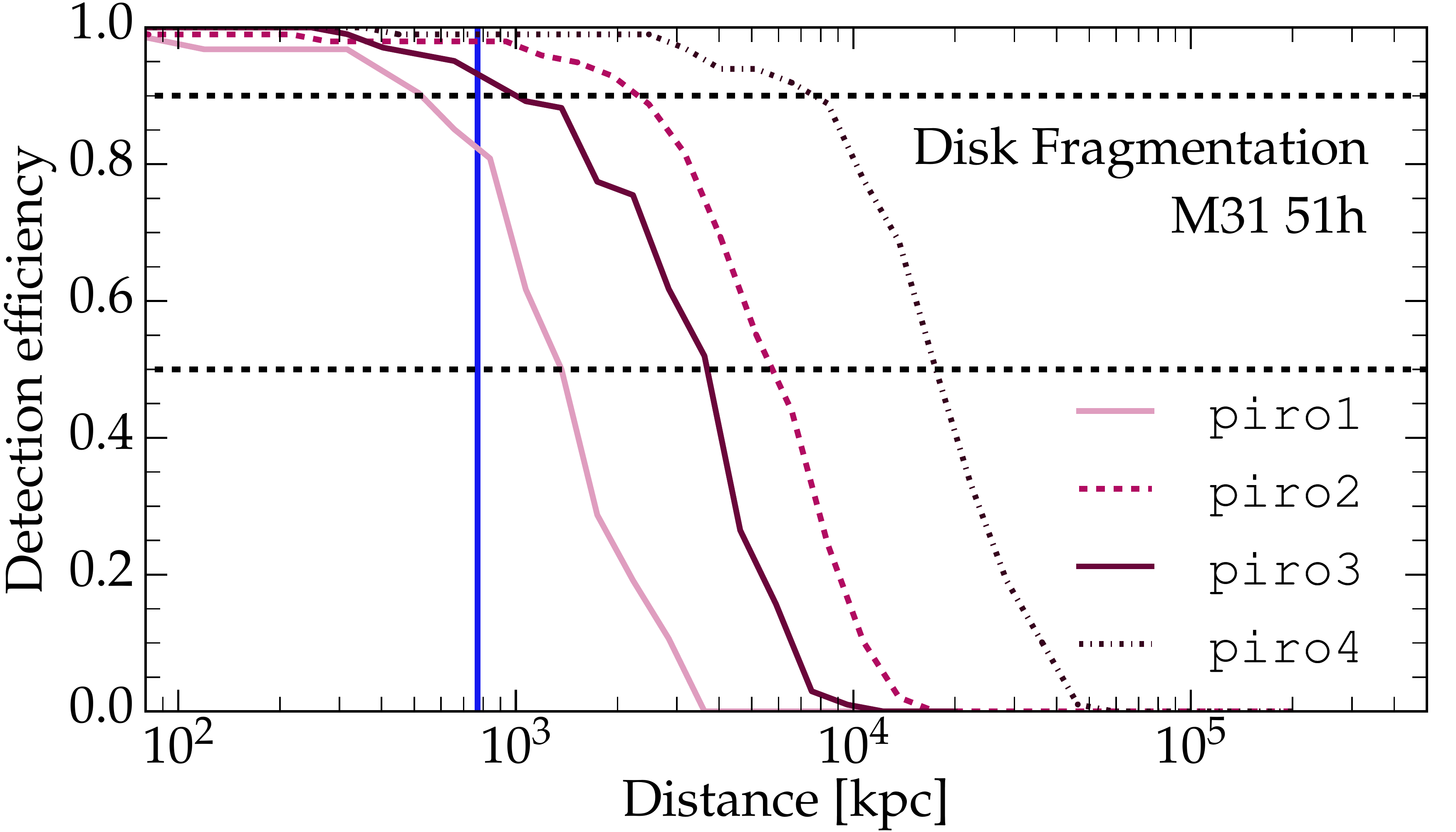}
\includegraphics[width=\columnwidth]{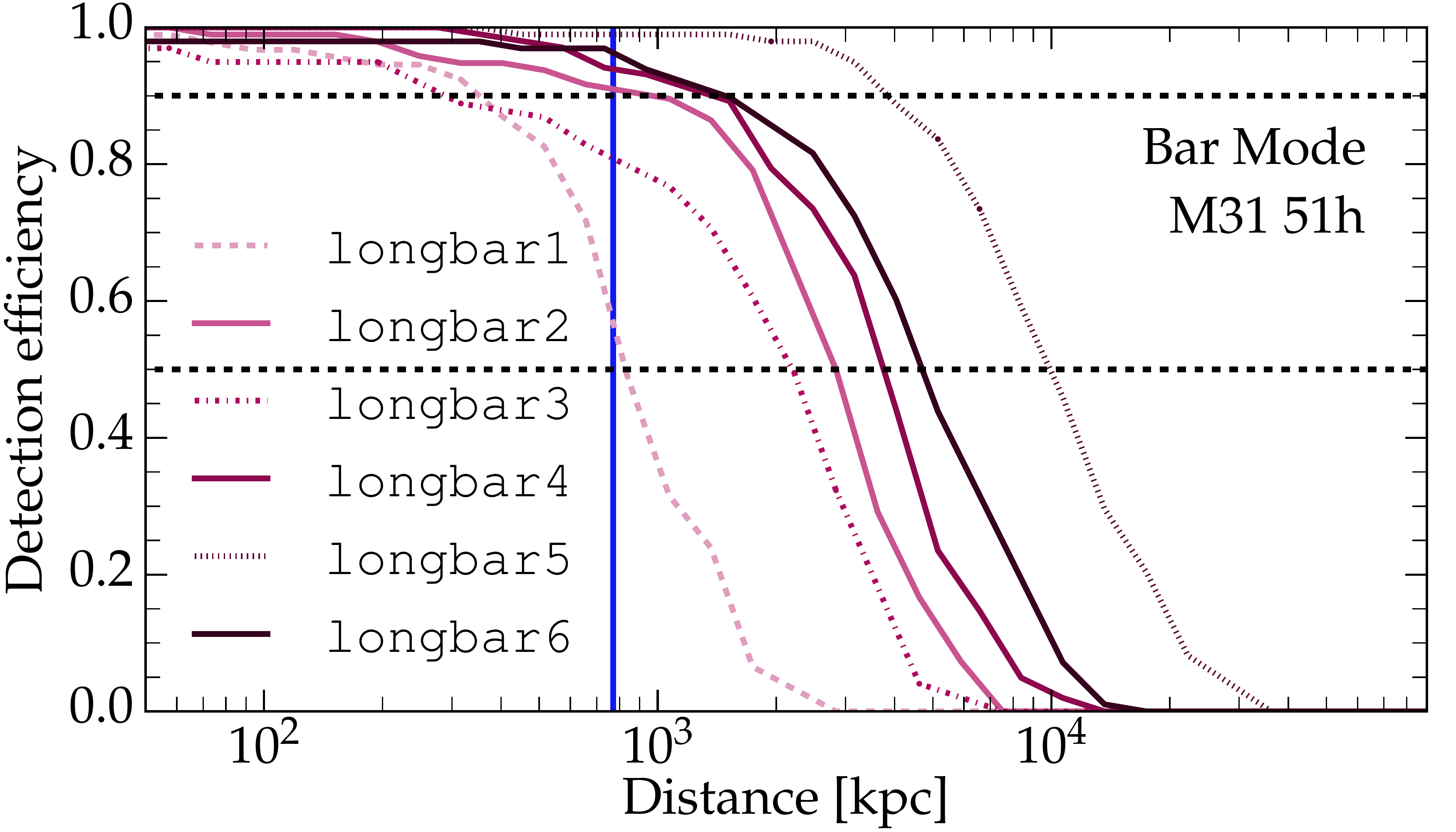}\\
\includegraphics[width=\columnwidth]{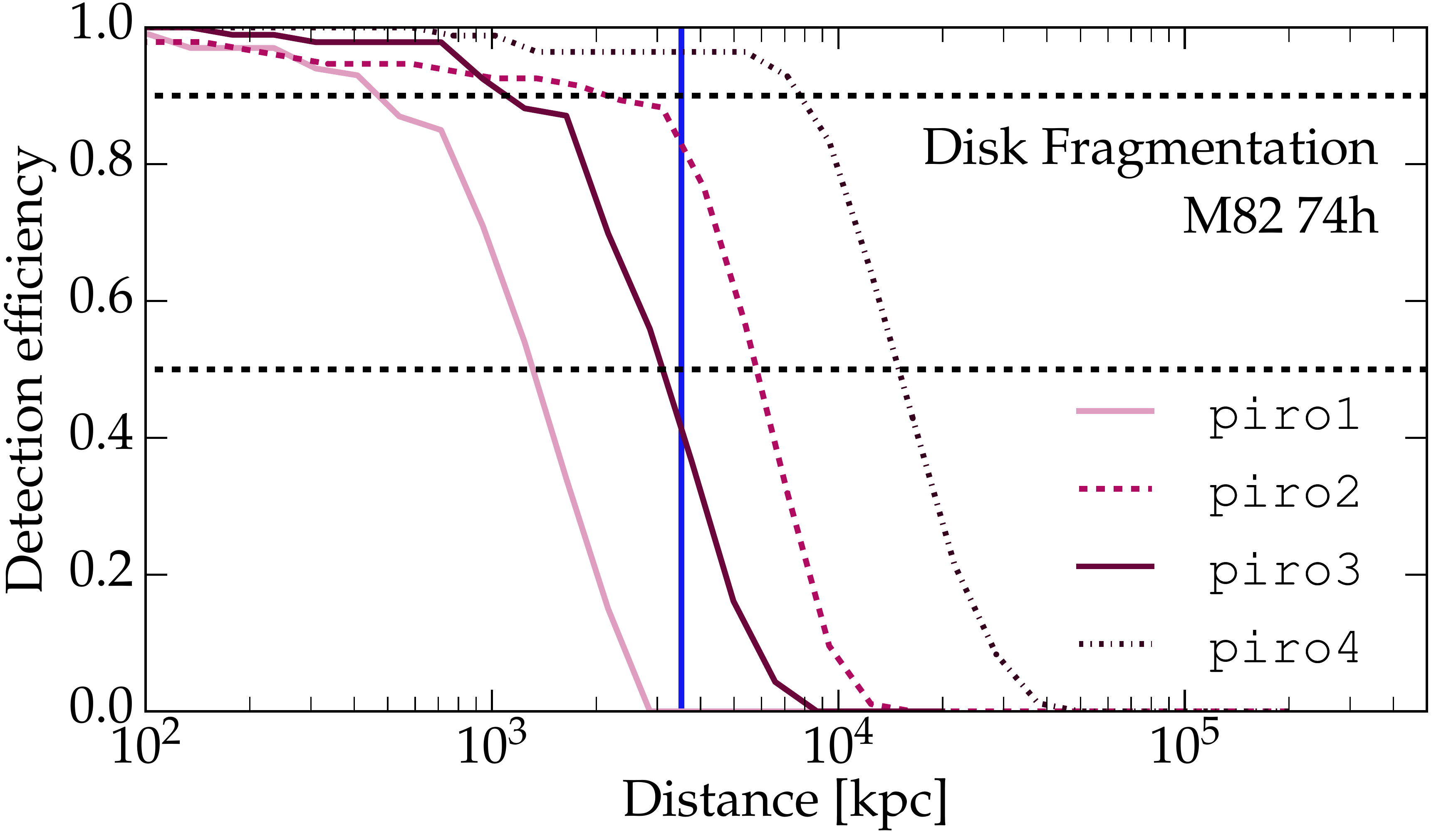}
\includegraphics[width=\columnwidth]{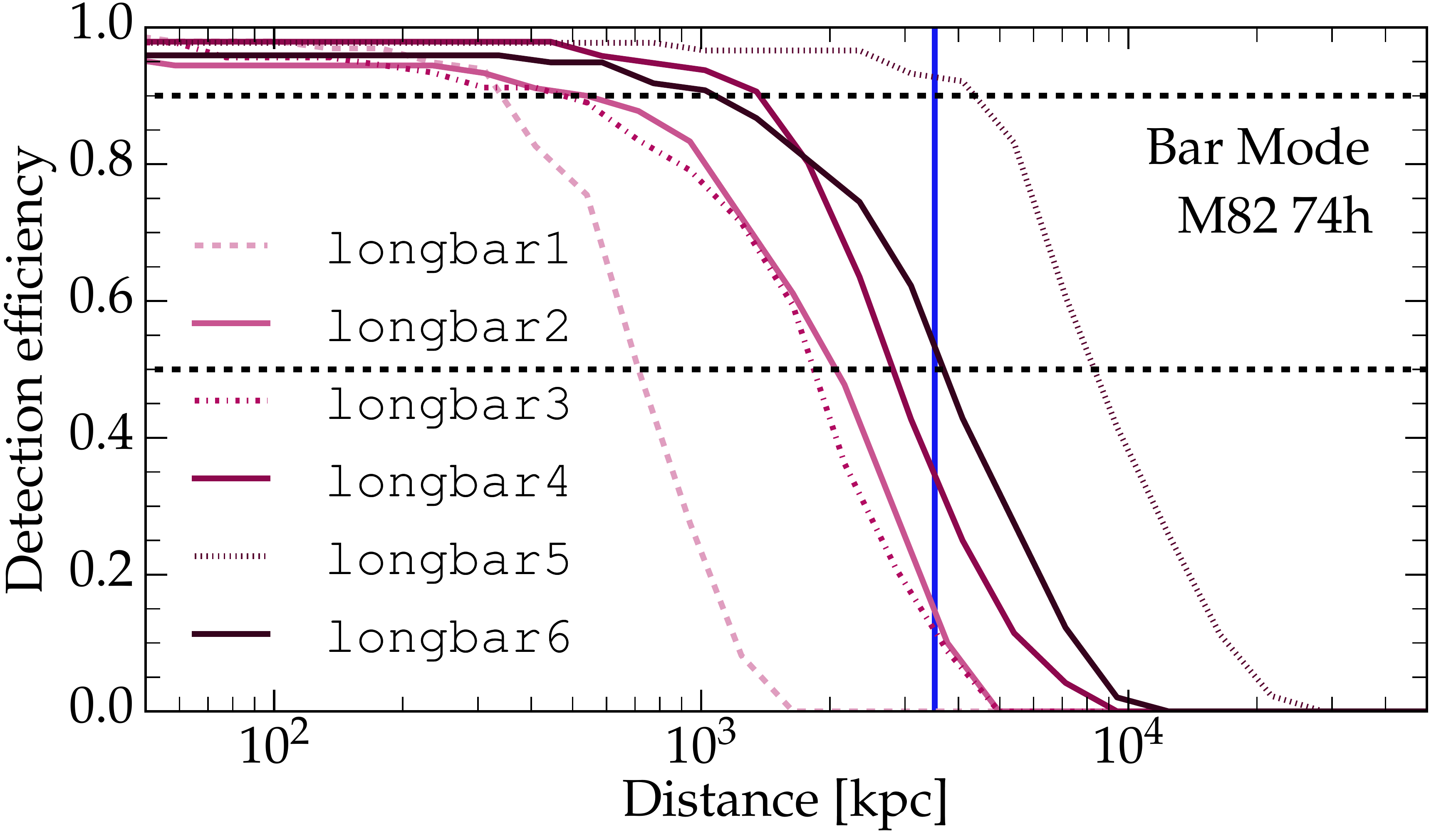}
\caption{The detection efficiency as a function of distance for the
phenomenological waveforms considered in this study, in the context of the 
on-source window astrophysically motivated by a type II SN
progenitor and the \texttt{HLV 2019} detector configuration. The top row is for
sources in M31 with an on-source window of 51 hours, and the bottom row is for
sources in M82 with a 74-hour on-source window. In each plot, 
$50\%$ and $90\%$ detection efficiency is marked with a
dashed black line, and the distance to the host galaxy is marked with a vertical
blue line.}
\label{fig:M31-M82_HLV2019B}
\end{figure*}

Figures~\ref{fig:M31-M82_HLV2019A} and \ref{fig:M31-M82_HLV2019B} present
the detection efficiency as a function of distance for the considered phenomenological 
extreme emission models, for CCSNe in the direction of M31 and M82 for the
\texttt{HLV 2019} detector network, using on-source windows motivated by type
Ibc and type II CCSNe, respectively.

\subsection{Sine-Gaussian waveforms}
\label{subsec:sgwaveform_results}
\begin{figure*}[!ht]
\includegraphics[width=\columnwidth]{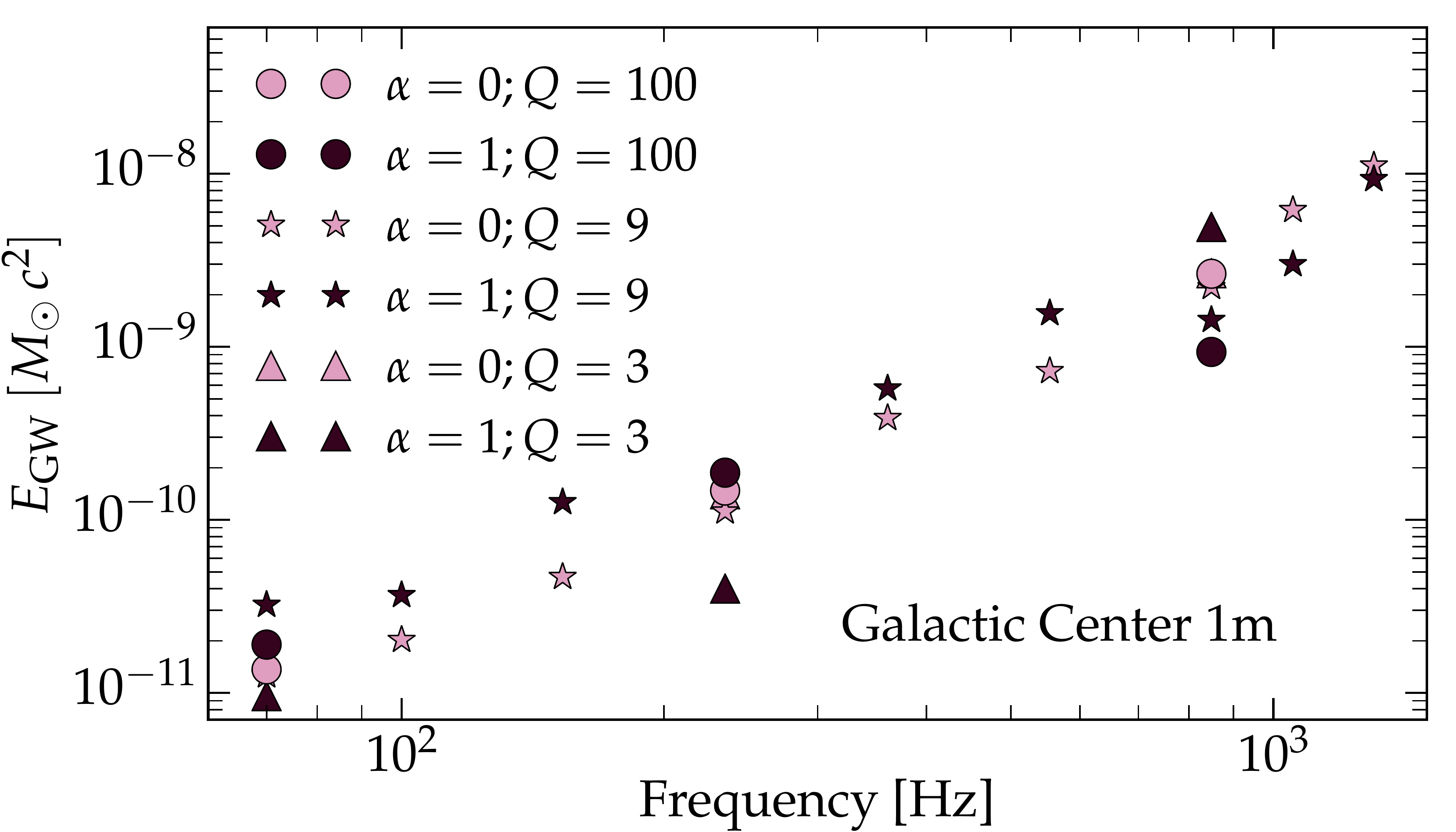}
\includegraphics[width=\columnwidth]{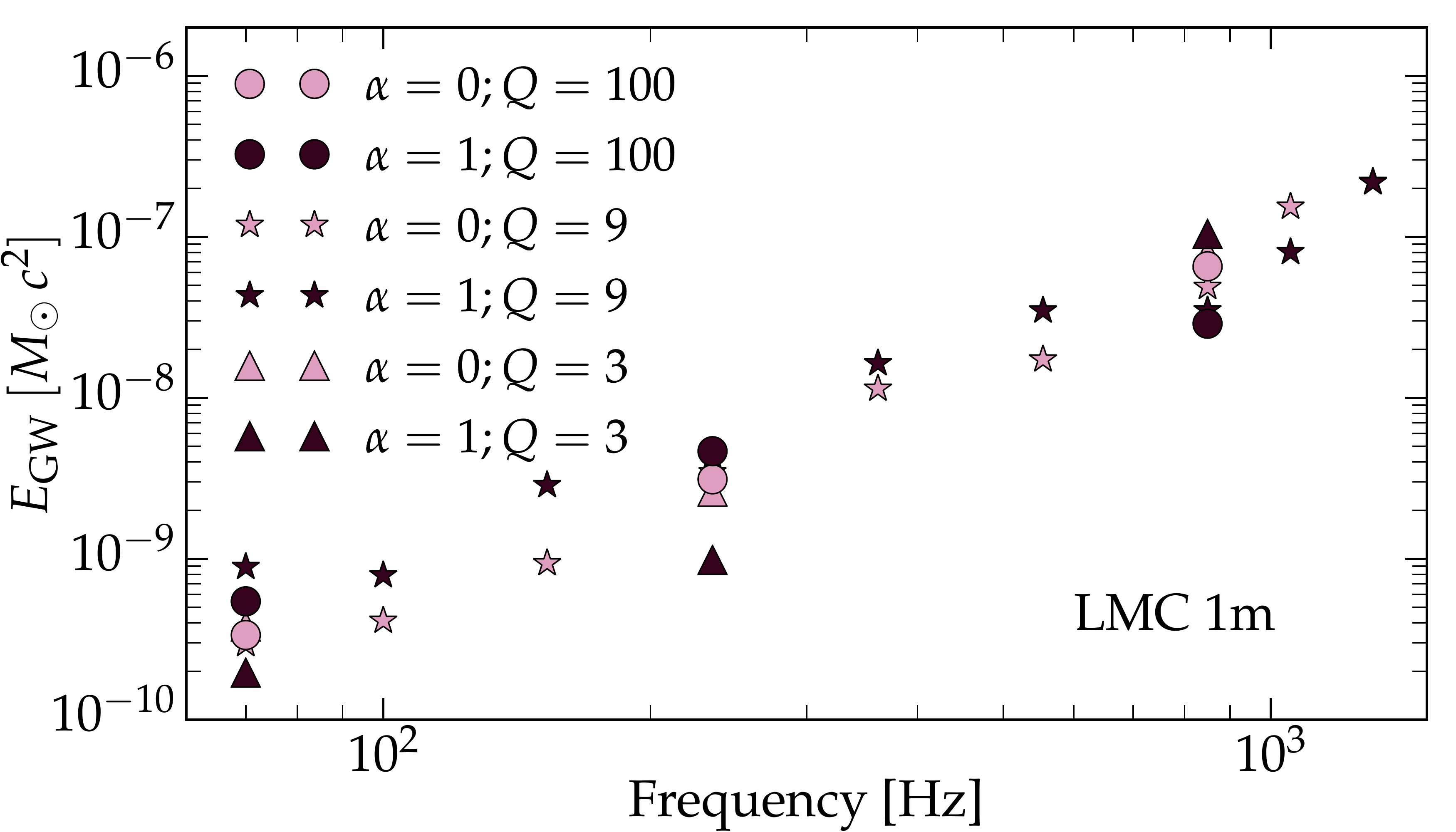}\\
\includegraphics[width=\columnwidth]{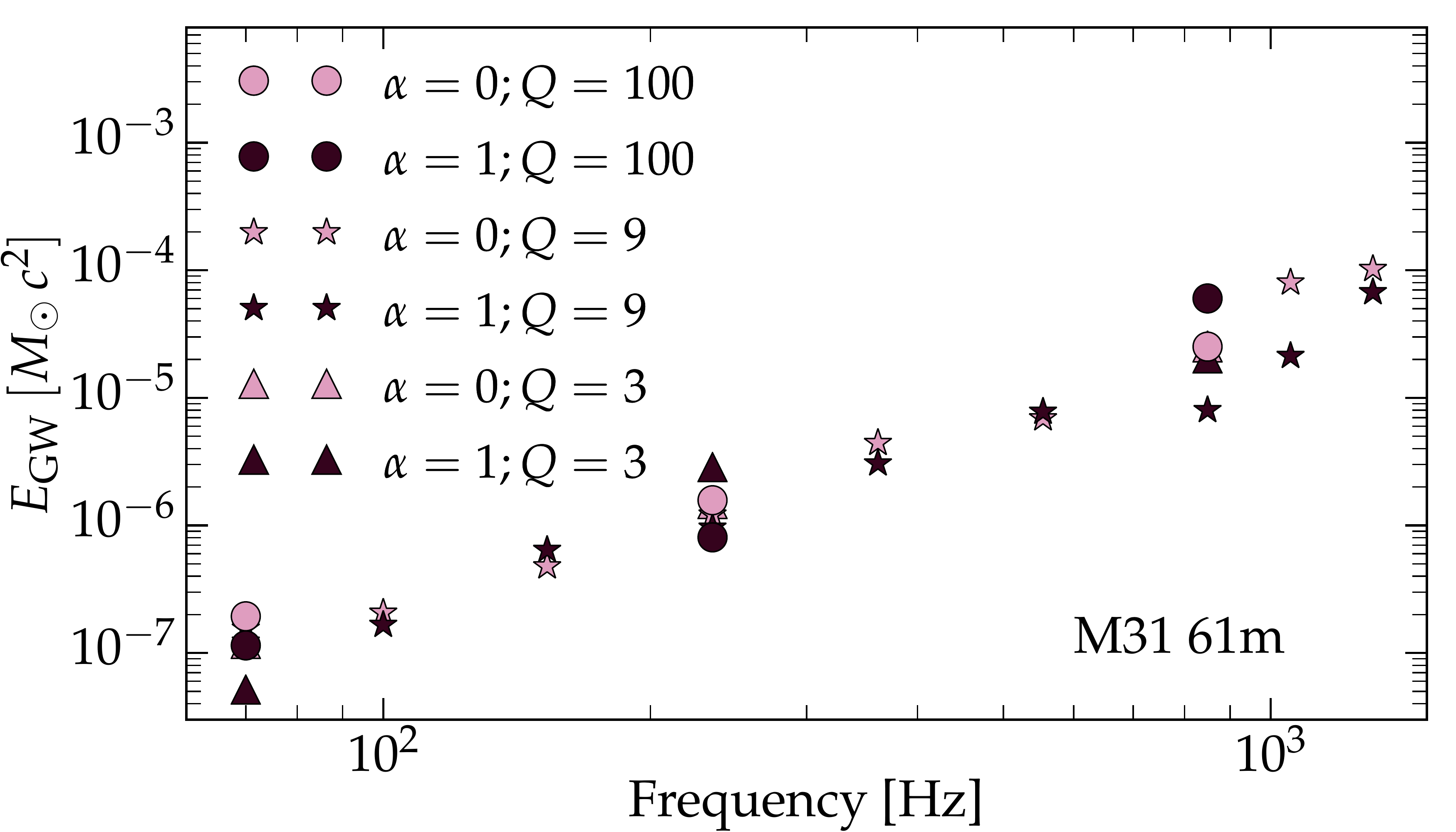}
\includegraphics[width=\columnwidth]{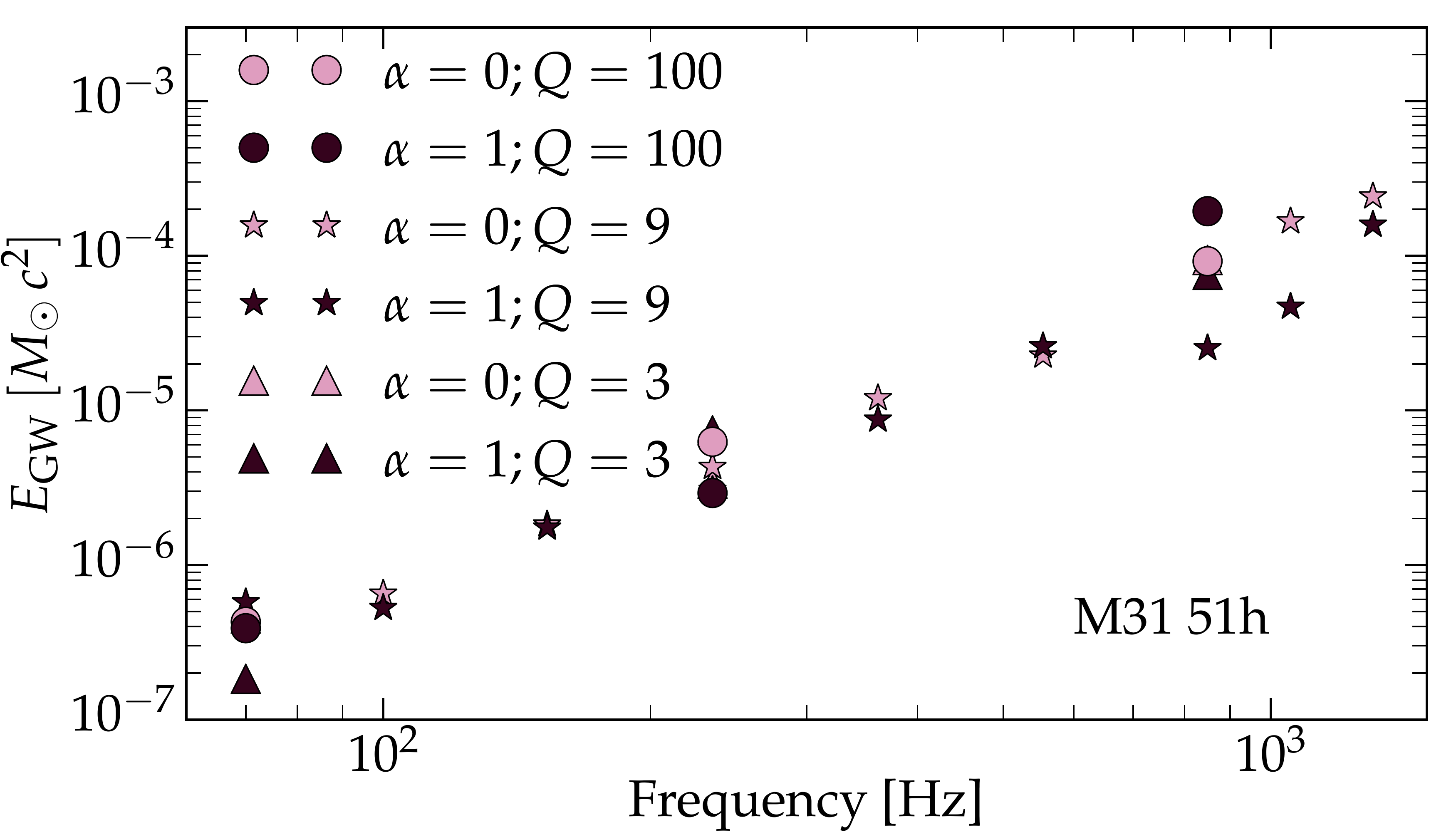}\\
\includegraphics[width=\columnwidth]{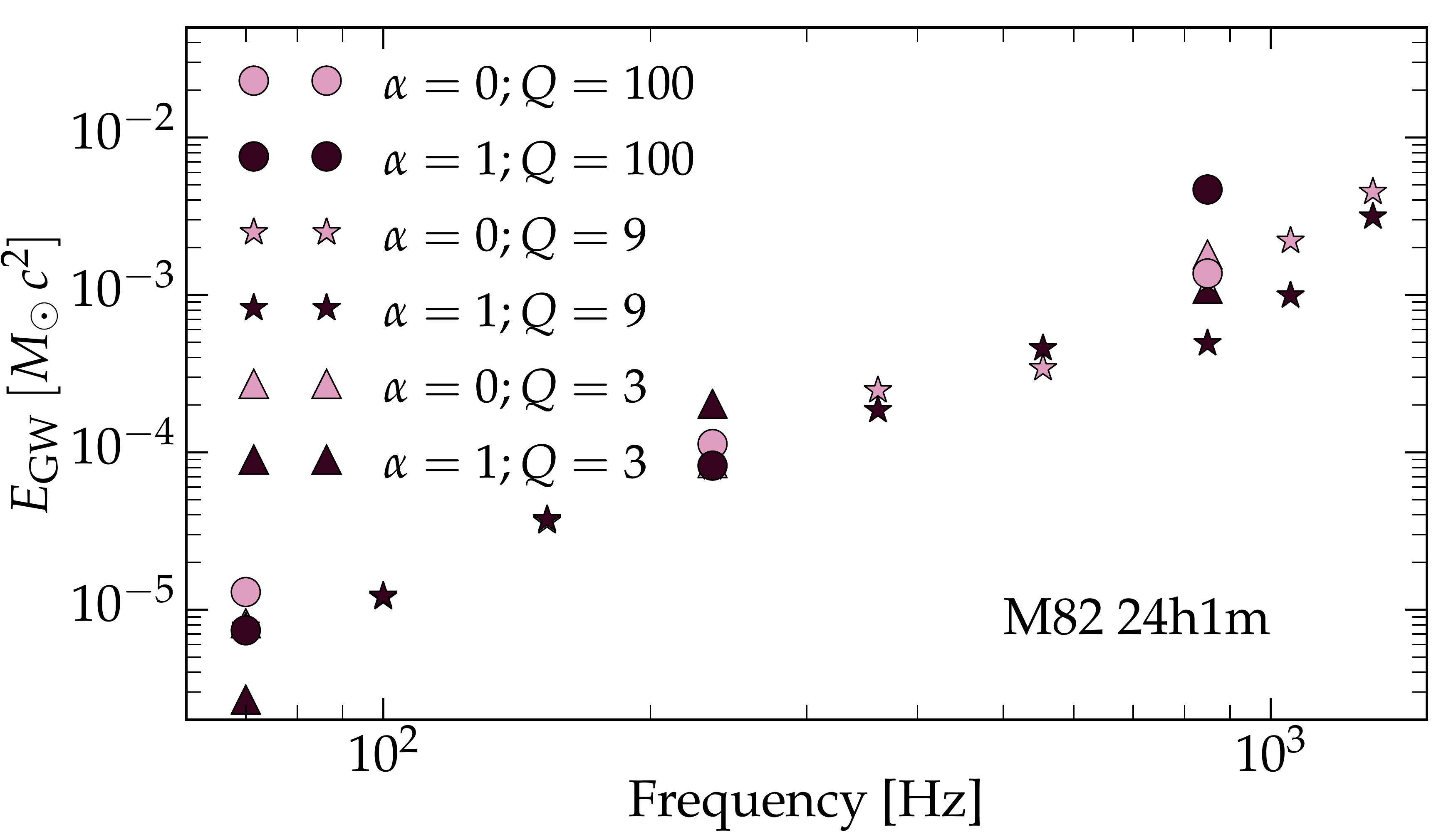}
\includegraphics[width=\columnwidth]{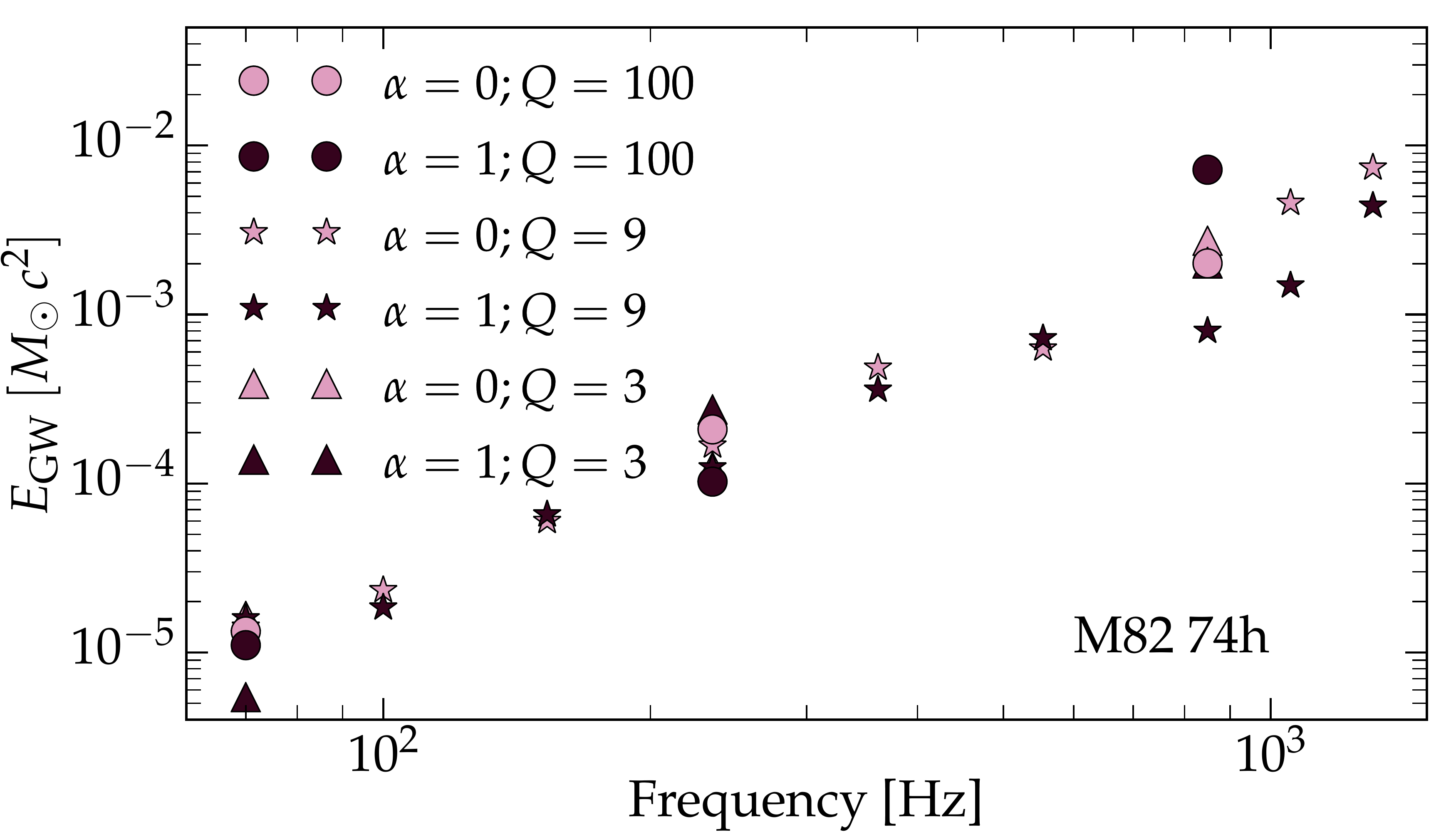}
\caption{The energy emitted in GW, $E^{50\%}_{\mathrm{GW}}$, required to attain the
root-sum-squared strain at $50\%$ detection efficiency, $h^{50\%}_{\mathrm{rss}}$, 
for the sine-Gaussian bursts considered in this study, in the 
context of the \texttt{HLV 2019} detector network. The top row is for sources directed 
toward the Galactic center (left) and the Large Magellanic Cloud (right), for both of
which a 1-minute on-source window is used. The middle row is for sources in M31, 
considering 61-minute and 51-hour on-source windows (left and right plots, respectively). 
The bottom row is for sources in M82, considering on-source windows of 24 hours and 
1 minute, and 74 hours (left and right plots, respectively). Distances of 
$10\,\mathrm{kpc}$, $50\,\mathrm{kpc}$, $0.77\,\mathrm{kpc}$, and 
$3.52\,\mathrm{Mpc}$ are used to compute $E^{50\%}_{\mathrm{GW}}$ for sources in
the galaxy, Large Magellanic Cloud, M31, and M82, respectively.} 
\label{fig:HLV2019_EGW}
\end{figure*}

The energy emitted in GW, $E^{50\%}_{\mathrm{GW}}$, required to attain the
root-sum-squared strain at $50\%$ detection efficiency, $h^{50\%}_{\mathrm{rss}}$, 
for the sine-Gaussian bursts considered is presented in
Fig.~\ref{fig:HLV2019_EGW}
for sources in the direction of the Galactic center, LMC, M31, and M82.

For the \emph{ad hoc} sine-Gaussian bursts considered, we use $E^{50\%}_{\mathrm{GW}}$ as the
figure of merit for detection.

For CCSNe in the direction of the Galactic center, we see that the typical
$E^{50\%}_{\mathrm{GW}}$ values are $\sim$\,(8-110)$\times10^{-10}\,\Msun$ 
for sine-Gaussian bursts with
central frequencies of $\sim$\,(554-1304)\,Hz, the typical frequencies of
emission for CCSNe, using a 60-second on-source window with the \texttt{HLV 2019}
detector network. For CCSNe in the direction of 
the LMC, we find $E^{50\%}_{\mathrm{GW}} \sim$\,(1-20)$\times10^{-8}\,\Msun$ in the
same frequency range. We
remind the reader that for the numerical waveforms considered, $E_{\mathrm{GW}}
\sim$\,(0.1-4000)$\times10^{-10}\,\Msun$. This is consistent, as
\texttt{X-Pipeline} is more sensitive to sine-Gaussian bursts, and we find that
only the more rapidly rotating models considered are detectable.

For CCSNe in the direction of M31, we find typical $E^{50\%}_{\mathrm{GW}}$ values of
$\sim$\,(7-100)$\times10^{-6}\,\Msun$ across the frequency range considered, using a 51
hour on-source window with the \texttt{HLV 2019} detector network. For CCSNe in the 
direction of M82, we find 
$E^{50\%}_{\mathrm{GW}}\sim$\,(3-60)$\times10^{-4}\,\Msun$ across the same 
frequency range. We remind the
reader that for the extreme phenomenological waveforms considered,
$E_{\mathrm{GW}} \sim$\,(2-600)$\times10^{-4}\,\Msun$. This is again consistent
with our previous results, as we find that all waveforms are detectable for
CCSNe in M31 with the \texttt{HLV 2019} detector network, but only the more
extreme cases are detectable out to M82.

\section{Discussion} 
\label{sec:conclusions}
The next galactic CCSN will be of great importance to the scientific community,
allowing observations of unprecedented accuracy via EM, GW, and neutrino
messengers. Using GW waveform predictions for core collapse from 
state-of-the-art numerical simulations, and phenomenological waveform models for
speculative extreme GW emission scenarios, we make the first comprehensive statements on
detection prospects for GWs from CCSNe in the Advanced detector era.

Given a known sky location, we outline a search procedure for GW bursts using
\texttt{X-Pipeline}, a coherent network analysis pipeline that searches for
excess power in time-frequency space, over some astrophysically motivated time 
period (or on-source window). The GW detector data is non-Gaussian,
nonstationary, and often contains loud noise transients. For this reason, it is
beneficial to minimize the on-source window to reduce the probability of
glitchiness or extreme Gaussian fluctuations being present in the detector data.

For CCSNe within $\sim 100\,\mathrm{kpc}$, the coincident neutrino signal will be detected,
allowing the time of core collapse to be determined to within a few tens of
milliseconds. Using an conservative asymmetric on-source window of $[-10,+50]$ seconds around
the start time of the neutrino signal, we consider hypothetical CCSNe in the
direction of the Galactic center and the LMC. We find that neutrino-driven CCSN
explosions, believed to account for $\sim99\%$ for CCSNe, will be detectable
within $2.4\,\mathrm{kpc}$, $3.5\,\mathrm{kpc}$, and $5.5\,\mathrm{kpc}$ in
2015, 2017, and 2019, respectively. Rapidly rotating CCSNe, however, will be 
detectable throughout the galaxy from $2017$,
and the most rapidly rotating model considered will be detectable out to the LMC
in $2019$. Rapidly rotating CCSNe with nonaxisymmetric rotational
instabilities will be detectable out to the LMC and beyond from 2015.

More distant CCSNe will not have coincident neutrino observations, and so the
on-source window must be derived using EM observations. Using recent studies of
light curves for type Ibc and type II CCSNe (see, e.g.
\cite{li:07,bersten:11,morozova:15}), we assume that, if the time of shock
breakout $t_{\mathrm{SB}}$ is observed, the time of core collapse can be 
localized to between 1 minute and 50 hours. Unfortunately, shock breakout 
is rarely observed, and an 
observation cadence time delay, $t_{\mathrm{obs}}$, between the last 
pre-CCSN and first post-CCSN images is introduced. Given this, we construct an
on-source window of $[-t_{\mathrm{SB}},t_{\mathrm{obs}}]$ about the time of the
last pre-CCSN image. Frequently observed galaxies, such as those for which the
CCSN rate is high, are likely to have CCSNe detected within one day of shock 
break-out. As such, we consider two observational scenarios where 
$t_{\mathrm{obs}} = 1\,\mathrm{hour}$ and $24\,\mathrm{hours}$ for 
hypothetical CCSNe in M31 and M82, respectively. 
In the context of EM observations of type Ibc CCSNe, we use on-source windows of
61 minutes and 24 hour 1 minute for CCSNe in M31 and M82, respectively.
Correspondingly for type II CCSNe, we use on-source windows of 51 hours and 74
hours for CCSNe in M31 and M82, respectively.
We find that most of the extreme GW
emission models considered are observable out to M31 with the
\texttt{HL 2015} detector network when using a 61-minute on-source window, while
all models are observable when using the 51-hour on-source window in 2019. Only
the most extreme emission models considered are observable out to M82 with 
the \texttt{HL 2015} detector network, but approximately half of the models
considered will be detectable out to M82 and beyond in 2019. This allows
us to either detect events associated with or exclude such extreme 
emission models for CCSNe in M31 and M82 with the
\texttt{HLV 2019} detector network.

In anticipation of unexpected GW emission from CCSNe, we additionally consider
sine-Gaussian bursts across the relevant frequency range for all observational
scenarios studied. We find, that the sensitivity of our search
method is comparable, if not slightly improved, to that found for the realistic
waveform models considered. This is to be expected as \texttt{X-Pipeline}, and
other clustering-based burst search algorithms, are most sensitive to short bursts 
of GW energy localized in frequency space. It should be noted, however, that
such simple waveform morphologies are more susceptible to being confused for
noise transients. As such, a more complicated waveform morphology, as found for
realistic GW predictions for CCSNe, can actually improve
detectability~\cite{kanner:15}.

Detection prospects for GWs from CCSNe can be improved by refining light curve
models for CCSNe, and increasing observation cadence, so as to reduce the
on-source window as derived from EM observations as much as possible. 
Improvement of stationarity and glitchiness of detector data, in 
addition to increasing the detector duty cycle, will improve detectability of
GWs from CCSNe. Further to this, more second-generation GW detectors such as
KAGRA and LIGO India will improve the overall sensitivity of the global GW
detector network and could potentially allow for neutrino-driven CCSN explosions 
to be observable throughout the Galaxy.

\begin{acknowledgments}
The authors thank Alan Weinstein, Peter Kalmus, Lucia Santamaria, 
Viktoriya Giryanskaya, Valeriu Predoi, Scott Coughlin, James Clark, 
Micha\l~W\k{a}s, Marek Szczepanczyk, Beverly Berger, and Jade Powell 
for many fruitful discussions that have benefitted this paper 
greatly. We thank the CCSN simulation community for making
their gravitational waveform predictions available for this
study. LIGO was constructed by the California Institute of Technology
and Massachusetts Institute of Technology with funding from the
National Science Foundation and operates under cooperative agreements
PHY-0107417 and PHY-0757058. C.D.O. is partially supported by National Science
Foundation Grants No. PHY-1404569 and No. CAREER PHY-1151197 and by the 
Sherman Fairchild Foundation. This paper has been assigned LIGO Document No.
LIGO-\ligodoc.
\end{acknowledgments}

\bibliography{methods_bib}

\end{document}